\documentclass{rmf-d}
\usepackage{nopageno,rmfbib,multicol,times,epsf,amsmath,amssymb,cite}
\usepackage[latin1]{inputenc}
\usepackage[]{caption2}
\usepackage{graphics}
\usepackage{graphicx}
\usepackage[hidelinks]{hyperref}
\hypersetup{colorlinks, citecolor=black, linkcolor=black ,urlcolor={blue}}
\usepackage{wrapfig, blindtext}
\usepackage{comment}
\usepackage{tabularx}
\usepackage{float}
\usepackage{dblfnote}
\newcommand{\Z}{\mathbb{Z}}

\newcommand{\R}{\mathbb{R}}
\newcommand{\C}{{\kern+.25em\sf{C}\kern-.45em\sf{{\small{I}}} \kern+.45em\kern-.25em}}
\newcommand{\aso}{\hat{=}}
\newcommand{\be}{\begin{equation}}
\newcommand{\ee}{\end{equation}}
\newcommand{\bea}{\begin{eqnarray}}
\newcommand{\eea}{\end{eqnarray}}
\newcommand{\nn}{\nonumber}

\newcommand{\la}{\langle}
\newcommand{\ra}{\rangle}

\def\rmfcornisa{\hfill\rmf\ {\bf }
\hfill }

\def\rmfcintilla{{\it Rev.\ Mex.\ Fis.\/} {\bf } }
\clearpage \rmfcaptionstyle 
\begin{document}
\title{The 3d O(4) model as an effective approach to the QCD phase diagram
\vspace{-6pt}}
\author{Edgar L\'{o}pez-Contreras}
\author{Jos\'{e} Antonio Garc\'{\i}a-Hern\'{a}ndez}
\author{El\'{\i}as Natanael Polanco-Eu\'{a}n}
\author{Wolfgang Bietenholz}
\address{ \ \\
  Instituto de Ciencias Nucleares, Universidad Nacional Aut\'{o}noma
de M\'{e}xico \\ A.P.\ 70-543, C.P.\ 04510 Ciudad de M\'{e}xico, Mexico}
\maketitle
\begin{abstract}
\vspace{1em}
The QCD phase diagram is one of the most prominent outstanding
puzzles within the Standard Model. Various experiments, which
aim at its exploration beyond small baryon density, are operating or
in preparation. From the theoretical side, this is an issue of
non-perturbative QCD, and therefore of lattice simulations.
However, a finite baryon density entails a technical problem
(known as the ``sign problem''), which has not been overcome
so far. Here we present a study of an effective theory, the
O(4) non-linear sigma model. It performs spontaneous symmetry
breaking with the same Lie group structure as 2-flavor QCD in
the chiral limit, which strongly suggests that they belong to the
same universality class. Since we are interested in high temperature,
we further assume dimensional reduction to the 3d O(4) model, which
implies topological sectors. As pointed out by Skyrme, Wilczek and
others, its topological charge takes the role of the baryon number.
Hence the baryon chemical potential $\mu_{B}$ appears as an imaginary vacuum
angle, which can be included in the lattice simulation without any
sign problem. We present numerical results for the critical line in
the chiral limit, and for the crossover in the presence of light quark
masses. Their shapes are compatible with other predictions, but up to
the value of about $\mu_{B} \approx 300~{\rm MeV}$
we do not find the notorious Critical Endpoint (CEP).

\vspace{1em}
\end{abstract}
\keys{QCD phase diagram, non-linear sigma model, lattice simulations \vspace{-4pt}}
\pacs{05.50.+q, 05.70.Fh, 12.38.-t, 12.39.Fe, 75.10.Hk  \vspace{-4pt}}
\begin{multicols}{2}

\section{The QCD phase diagram}

\begin{figure}[H]
\centering
\includegraphics[scale=0.85]{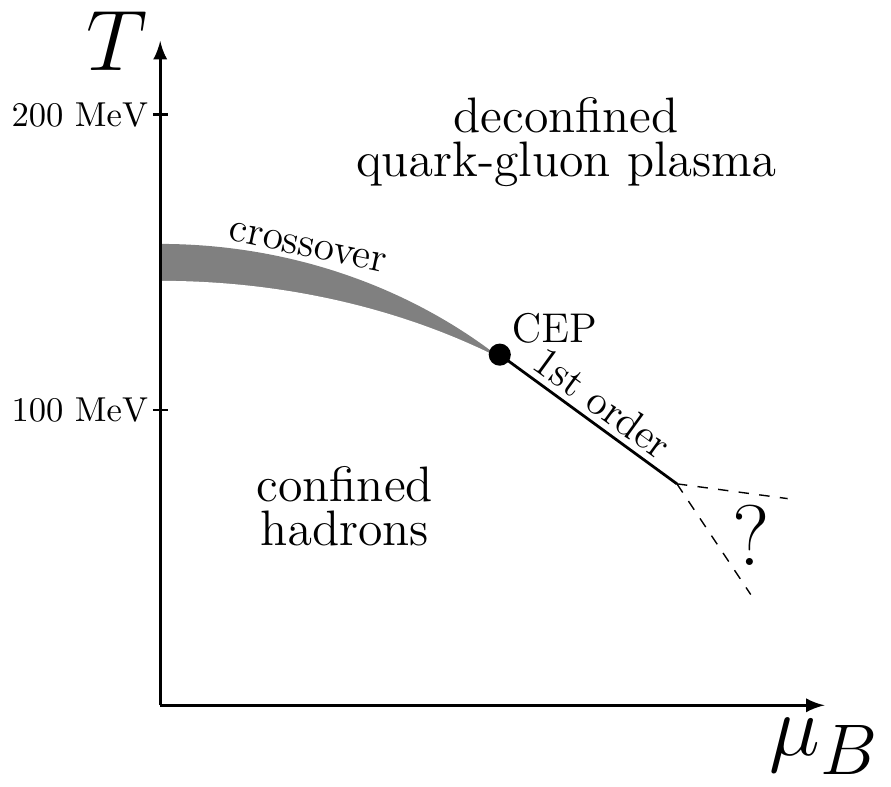}
\caption{Symbolic illustration of the expected QCD phase diagram.}
\vspace*{-0.2cm}
\label{phasedia1}
\end{figure}

Beyond low baryon density, the QCD phase diagram is still
{\it terra incognita}, both theoretically and experimentally
(we assume the validity of QCD as the correct theory of the
strong interaction to persist). It can be parameterized by the
inclusion of a baryonic chemical potential $\mu_{B}$, which
characterizes the density of the net baryon number $B - \bar B$,
as sketched in Fig.\ \ref{phasedia1}.

It is often a good approximation to assume the light quark masses
to be degenerate; we denote this mass as $m_{q}:= m_{u} = m_{d}$.
In this setting, at $\mu_{B} = 0$, lattice QCD simulations provide
the following results:
\begin{itemize}

\item
  In the chiral limit of $u$ and $d$ quarks, $m_{q}=0$, one obtains
  a second order phase transition between the confined (hadronic) and
  deconfined phase (quark-gluon plasma).
  With the $s$-quark included, the critical temperature amounts
  to $T_{\rm c} \simeq 132~{\rm MeV}$ \cite{Ding19}. If we still add the
  $c$-quark (with phenomenological values of $m_{s}$ and $m_{c}$),
  the transition turns into a crossover, but its temperature hardly
  changes; one obtains the pseudo-critical temperature
  $T_{\rm pc} \simeq 134~{\rm MeV}$ \cite{KLT21}.

\item For a realistic $m_{q} >0$, and $2$ or $2+1$ flavors, one obtains
  a crossover as well. In the latter scenario, the pseudo-critical
  temperature is somewhat higher, $T_{\rm pc} \simeq 155~{\rm MeV}$
  \cite{BazavovBhattacharya}. This is consistent with the
  experimentally measured freeze-out temperature of the quark-gluon
  plasma.
  
\end{itemize}

This phase diagram is of interest {\it e.g.}\ for our understanding
of the early Universe and of neutron stars.
Several experiments are operating with the goal of exploring
the nuclear phase diagram, at facilities like the Super Proton
Synchrotron (SPS), the Relativistic Heavy Ion Collider (RHIC) and the
Large Hadron Collider (LHC). Others are in preparation, we mention
the Facility for Antiproton and Ion Research (FAIR), and in
particular the Multi-Purpose Detector at the Nuclotron-based Ion Collider
fAcility (MPD-NICA), which is under construction at JINR in Dubna, Russia
\cite{firstphysics}, with the participation of the Mexican group MexNICA.
It plans to collide heavy ions, such a bismuth nuclei, at energies of
$4$ to $11~{\rm GeV}$ per nucleon, which is suitable to attain a high
baryon number density, and to access the region where one expects the
Critical Endpoint (CEP), {\it i.e.}\ the point in the phase diagram
where the crossover turns into a first order phase transition,
cf.\ Fig.\ \ref{phasedia1}.

However, the location --- and even the existence --- of the CEP is
uncertain. If it exists, one speculates about a rich phase structure at
even higher $\mu_{B}$, including for instance a color superconducting phase.

From the theoretical side, this is an issue of non-perturbative QCD,
and therefore of lattice simulations, which did provide the aforementioned
values of $T_{\rm c}$ and $T_{\rm pc}$. It deals with the QCD formulation in
Euclidean space-time, which is justified for equilibrium observables.
One further assumes a discrete lattice structure, which implements
an UV regularization. The quark fields $\psi_{x}$ are formulated on the
lattice sites $x$, and the gluon fields $U_{x,\mu}$ on the links
connecting them ($\mu$ specifies the direction). It is
profitable to use {\em compact} link variables in the gauge group
(not in the algebra), $U_{x,\mu} \in {\rm SU}(3)$, which avoids the
need of gauge fixing. In analogy to Statistical Mechanics, one
introduces the partition function in the functional integral formalism,
\bea
Z &=& \int D \bar \psi D \psi D U e^{-S_{\rm quark}[\bar \psi, \psi, U]
  - S_{\rm gauge}[U]} \nn \\
&=& \int DU \ {\rm det} M[U] \ e^{- S_{\rm gauge}[U]} \ .
\eea
The factor ${\rm det} M[U]$ is the fermion determinant,
which captures in particular
the sea quark contributions. Its numerical computation is tedious,
but one does not need to deal explicitly with the Grassmann-valued fields
$\bar \psi$, $\psi$. Thus we obtain expectation values of observables,
in particular $n$-point functions, as
\be
\la \dots \ra = \frac{1}{Z} \int DU ( \dots )
    \ {\rm det} M[U] \ e^{- S_{\rm gauge}[U]} \ .
\ee
The method consists of generating a large set of gauge configurations
$[U]$ with the probability distribution
\be
p[U] = \frac{1}{Z} \ {\rm det} M[U] \ e^{- S_{\rm gauge}[U]} \ ,
\ee
which enables the numerical measurement of $\la \dots \ra$. Here, we
assume the Euclidean action
$S_{\rm QCD} =- \ln {\rm det} M[U]  + S_{\rm gauge}[U]$
to be real positive, and we see that the Euclidean metrics is vital.

This method provides results with statistical errors (due to the finite
set of configurations), and systematic errors (we need to extrapolate
to the continuum and to infinite volume), but they are controlled
and additional simulations reduce them. This approach is fully
{\em non-perturbative:} a strong coupling like $\alpha_{\rm s} = {\cal O}(1)$
does not cause any problem.

The temperature $T$ is given by the inverse extent in Euclidean time,
which should be much shorter than the spatial directions to obtain results
at finite $T$. This is how $T_{\rm c}$ and $T_{\rm pc}$ were obtained.

However, adding a chemical potential $\mu_{B} > 0$ leads to a serious
difficulty known as the ``sign problem''. We can interpret $\mu_{B}$
as the energy, which is required for adding one more baryon.
It multiplies a real Lagrangian term in Minkowski space, but this term
becomes imaginary under Wick rotation. With this term,
${\rm det} M$, and therefore also the Euclidean
action $S_{\rm QCD}$, is complex, so $\frac{1}{Z} \exp (-S_{\rm QCD})$
does not represent a probability anymore.
In this case (and similarly in the presence of a $\theta$-term),
the standard technique that we sketched above does not apply.

Numerous attempts have been studied to overcome the sign problem,
but there is no breakthrough so far. For comprehensive reviews,
we refer for instance to Refs.\ \cite{Philippe}.
\begin{itemize}

\item The straight approach is simulating with probability
  $p \propto \exp (- {\rm Re} \, S_{\rm QCD})$, and including
  the complex phase {\it a posteriori} by re-weighting.
  This is correct in principle, but it leads to excessive
  cancellations, such that a precise result requires huge
  statistics. With stable statistical errors, the requested
  statistics grows exponentially with the volume, which often
  makes this approach hopeless.

\item The complex Langevin algorithm can handle and update a complex
  action, but the link variables leave the gauge group SU(3).

\item Some collaborations simulate at imaginary chemical potential,
  $\mu_{B}^{2} <0$, and try to extrapolate to $\mu_{B}^{2} >0$.

\item At $\mu_{B}=0$ it is possible to compute some coefficients of
  the Taylor series of the crossover curve, which extends to $\mu_{B} >0$.
  
\end{itemize}
Unfortunately none of these approaches is really conclusive regarding
the search for the CEP.

Quantum computing offers some hope: it would 
allow us to directly deal with $S_{\rm QCD} \in \C$. This is under
intense investigation in toy models, but not yet applicable to QCD.
We mention one example, which refers to analogue quantum computing,
with Mexican participation \cite{qcomp}: if one traps suitable,
ultra-cold alkaline-earth atoms in the nodes of a 2d optical lattice,
the nuclear spins represent an SU(3) field, which may perform
Spontaneous Symmetry Breaking (SSB), ${\rm SU}(3) \to {\rm U}(2)$.
Then the low-energy effective action of the
Nambu-Goldstone bosons just corresponds to the 2d $\C$P(2) model,
which could be quantum simulated in this manner, and which bears
a number of similarities with QCD (asymptotic freedom, topology,
a dynamically generated mass gap).

In the absence of conclusive QCD results, one derives conjectures about
the QCD phase diagram from related models. Many such models have been
studied.
Examples, and corresponding references, 
include the Nambu-Jona-Lasinio model \cite{NJL}, and more
specifically the Polyakov-Nambu-Jona-Lasinio model \cite{PNJL},
the linear $\sigma$-model \cite{LSM},
holographic approaches to QCD \cite{holography},
the Polyakov quark meson model \cite{quarkmeson},
as well as methods like
the Dyson-Schwinger equation \cite{DSE},
the mean-field approximation \cite{MFA}
and finite-size scaling \cite{FSS}.

As a new approach, here we focus on the 3d O(4) non-linear
$\sigma$-model, with an imaginary $\theta$-term.

\section{The 3d O(4) model as an effective theory}

\subsection{2-flavor QCD}

Two quark flavors are very light compared to the intrinsic scale of
QCD, $m_{u} \simeq m_{d} \ll \Lambda_{\rm QCD} \approx 300~{\rm MeV}$,
hence the chiral limit $m_{q} := m_{u}=m_{d} = 0$ is often a good
approximation. (For instance, the nucleon mass is only modified by
a few percent, which shows that the mass of macroscopic objects
is mostly due to the gluon energy, and only to a minor part due to
the Higgs mechanism.)
In this limit, the left- and right-handed quarks decouple,
\be
   {\cal L}_{\rm quark} = (\bar u, \bar d)_{\rm L} \gamma_{\mu} D_{\mu}
   \left( \begin{array}{c} u \\ d \end{array} \right)_{\rm L}
   + (\bar u, \bar d)_{\rm R} \gamma_{\mu} D_{\mu}
   \left( \begin{array}{c} u \\ d \end{array} \right)_{\rm R} \ , \nn
\ee
so the corresponding quark doublets can be transformed
independently, and the QCD Lagrangian has the global
symmetry
\be
   {\rm U}(2)_{\rm L} \otimes {\rm U}(2)_{\rm R} =
   {\rm SU}(2)_{\rm L} \otimes {\rm SU}(2)_{\rm R}
\otimes {\rm U}(1)_{\rm L=R} \otimes {\rm U}(1)_{\rm axial} \ . \nn
\ee
The ${\rm U}(1)_{\rm L=R}$ symmetry assures the fermion number conservation,
while the axial symmetry ${\rm U}(1)_{\rm axial}$ is anomalous (explicitly
broken under quantization). At $T < T_{\rm c}$ the remaining chiral
flavor symmetry undergoes SSB,
\be
   {\rm SU}(2)_{\rm L} \otimes {\rm SU}(2)_{\rm R} \longrightarrow
   {\rm SU}(2)_{\rm L=R} \ ,
\ee
which --- according to Goldstone's Theorem --- generates 3 Nambu-Goldstone
bosons. If we add small quark masses to the $u$- and $d$-quark, they
become massive, because the symmetry breaking has a (small) explicit
component, and these quasi-Nambu-Goldstone are identified with the pions.

\subsection{The O(4) model as an effective theory}

We proceed to the O(4) non-linear $\sigma$-model as an effective theory
with an equivalent SSB pattern. Its action reads
\be  \label{actO4}
S [\vec e \, ] = \int d^4x \left[ \frac{F_{\pi}^{2}}{2}
  \partial_{\mu} \vec e(x) \cdot \partial_{\mu} \vec e(x)
  - \vec h \cdot \vec e(x) \right] \ ,
\ee
with $\vec e(x) \in S^{3}$, and $\vec h$ is an external ``magnetic field''
(or ``ordering field''). According to Chiral Perturbation Theory,
$F_{\pi} \simeq 92.4~{\rm MeV}$ is the pion decay constant.

At $\vec h = \vec 0$ the action has a global O(4) symmetry, which
can break spontaneously to O(3) (``spontaneous magnetization'').
$\vec h \neq \vec 0$ adds some explicit symmetry breaking, like
the (degenerate) quark mass $m_{q} >0$. The symmetry groups with or
without SSB, or quasi-SSB, are locally isomorphic,
\be
\{ \ {\rm SU}(2)_{\rm L} \otimes {\rm SU}(2)_{\rm R} \ \aso \ {\rm O}(4) \ \}
\longrightarrow \{ \ {\rm SU}(2)_{\rm L=R} \ \aso \ {\rm O}(3) \ \} \ . \nn
\ee
The SSB pattern and the space-time dimension usually determine
the universality class at criticality, so we have a strong reason
to assume the O(4) model to belong to the same universality class as
2-flavor QCD, cf.\ Refs.\ \cite{chiralPT}.

In the broken phase, it can be regarded as an effective pion
model, as in Chiral Perturbation Theory, since the field is
defined in the SSB coset space,
$\vec e \in S^{3} = {\rm O}(4)/{\rm O}(3)$. Hence we deal with
a meson field, so how can we address the baryon number?

Unlike Chiral Perturbation Theory, we are interested in high
$T = 1/\beta$. We assume it to be high enough for dimensional
reduction to be a good approximation, {\it i.e.}\ we assume
the dominant configurations $[\vec e \, ]$ to be (nearly) constant
in the (short and periodic) Euclidean time direction.\footnote{This
  assumption can be questioned, {\it i.e.}\ one may wonder whether
  $1/T_{\rm pc} \simeq 1.3 ~{\rm fm}$ is small enough to justify this
  simplification. One can further object that at high-$T$ heavier
  quark flavors are not negligible, but we cannot include them in
  O($N$) model effective theories \cite{WB}. Still, we are confident
  that our assumptions are sensible approximations.}
This reduces the temporal integral in the action (\ref{actO4}) to a constant,
$\int_{0}^{\beta} dt_{\rm E} \approx \beta$, and we obtain
(in a spatial volume $V$)
\bea
S[\vec e\, ] &=& \beta \int_{V} d^{3}x \ \left[
  \frac{F_{\pi}^{2}}{2} \partial_{i} \vec e(x) \cdot
  \partial_{i} \vec e(x) - \vec h \cdot \vec e(x) \right] \nn \\
  & = & \beta H[\vec e\, ] \ .
\eea
Thus we arrive at the 3d O(4) model, with periodic boundary conditions,
which has topological sectors, due to $\pi_{3}(S^{3}) = \Z$. The topological
charge $Q \in \Z$ represents the winding number of a configuration
$[\vec e\,]$ on $S^{3}$, which is invariant under (almost all) small
deformations of $[\vec e\,]$.

Skyrme and others noticed that the topological charge $Q$ of the
effective theory corresponds to the baryon number $B$ \cite{Skyrme}.
This identification can be derived from anomaly matching. Thus the meson
field does account for the baryon number, by means of topological windings.
Hence in the effective theory, the baryonic chemical potential $\mu_{B}$
takes the role of an imaginary vacuum-angle $\theta$,
\be
H[\vec e\, ] = \dots - \mu_{B} Q[\vec e\, ] \in \R \ .
\ee
We see that it can be incorporated in the effective theory
{\em without} any sign problem.

\subsection{The 3d O(4) model on the lattice}

In order to simulate the 3d O(4) model we need to formulate it on
the lattice. We choose the standard formulation on a cubic lattice,
and use lattice units ({\it i.e.}\ we set the lattice 
spacing to 1). The derivatives are replaced by nearest-neighbor
differences,
$$
\tfrac{1}{2} \partial_{i} \vec e(x) \cdot \partial_{i} \vec e(x)
\to  \tfrac{1}{2} (\vec e_{x + \hat i} - \vec e_{x} )^{2}
= 1 - \vec e_{x} \cdot \vec e_{x + \hat i} \ ,
$$
\be
S_{\rm lat}[\vec e\, ] = -\beta_{\rm lat} \bigg(
\sum_{\la xy \ra} \vec e_{x} \cdot \vec e_{y} + \vec h_{\rm lat} \cdot
\sum_{x} \vec e_{x} + \mu_{B,{\rm lat}} Q[\vec e\, ] \bigg) \ , \nn
\ee
where $\hat i$ is a unit vector in $i$-direction, and
$\la xy \ra$ are the nearest-neighbor lattice sites
(the constant $1$ can be dropped).

We formulate the topological charge of a lattice configuration
with a geometric definition. Thus we generalize the formulation
of Ref.\ \cite{BergLuscher}, which guarantees $Q[\vec e\,] \in \Z$
for all configurations (up to a subset of measure zero).

\begin{figure}[H]
\centering
\hspace*{2mm}
\includegraphics[scale=0.19]{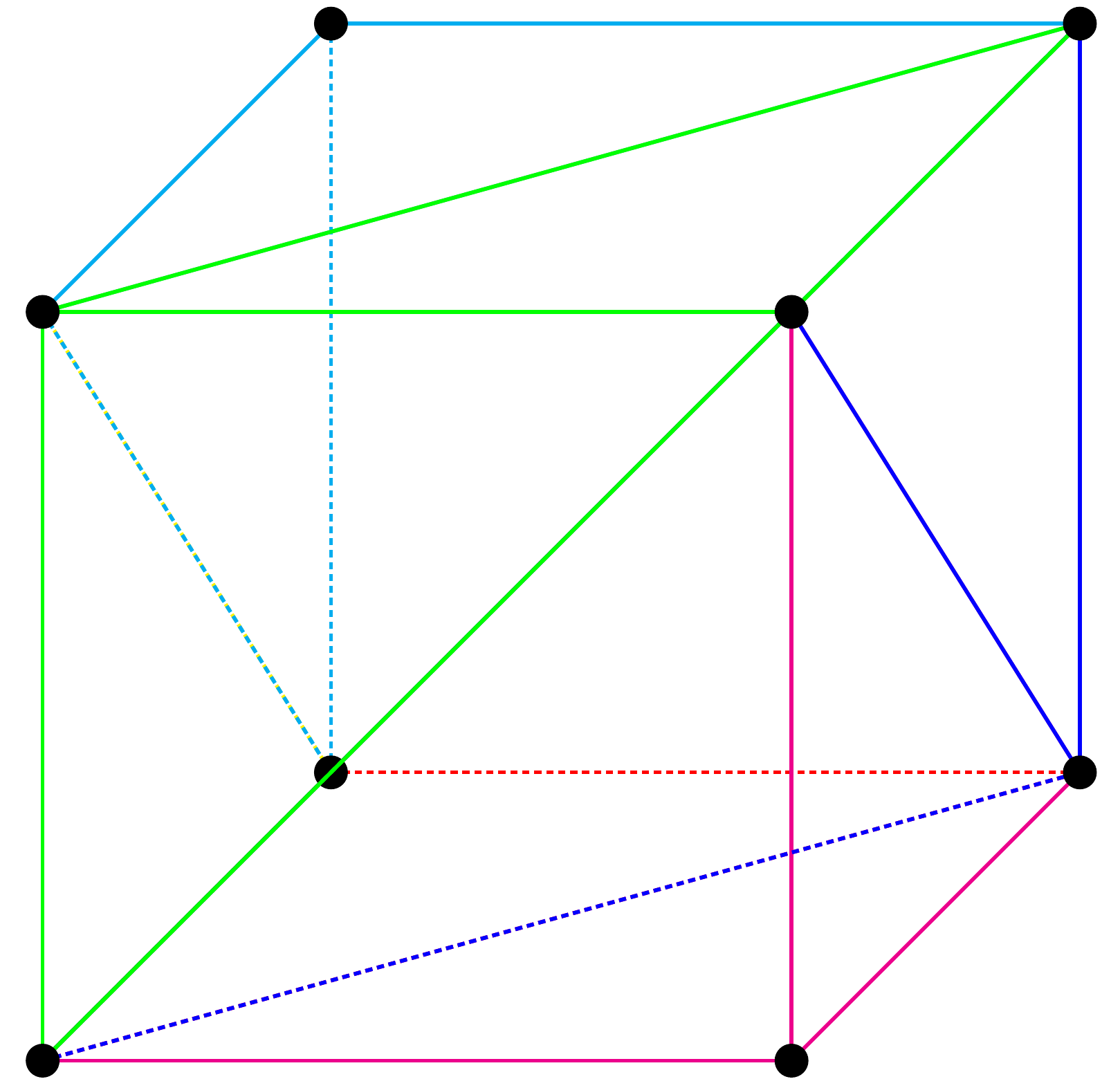}
\hspace*{7mm}  
\includegraphics[scale=0.38]{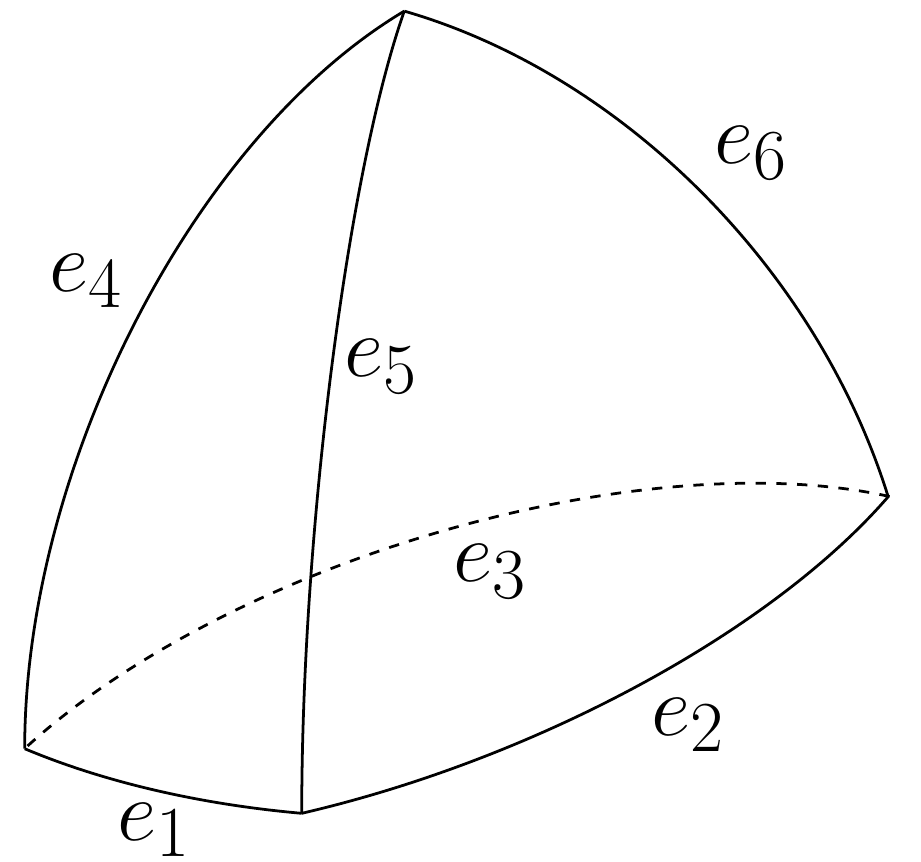}
\vspace*{2mm}
\caption{Left: Division of a lattice unit cube into 6 tetrahedra.
  Right: Symbolic illustration of a spherical tetrahedron.}
\label{tetra}
\end{figure}

To be explicit, we split the lattice unit cubes into 6 tetrahedra, as shown
in Fig.\ \ref{tetra} (left). The 4 spins at the vertices of one tetrahedron
--- we call them $(\vec e_{w}, \vec e_{x}, \vec e_{y}, \vec e_{z})$ ---
span a {\em spherical tetrahedron} on $S^{3}$, as symbolically sketched
in Fig.\ \ref{tetra} (right): its edges $e_{1} , \dots , e_{6}$ are
geodesics in $S^3$.
The topological density of a tetrahedron is given by the oriented,
normalized volume of its corresponding spherical tetrahedron,
$V_{w,x,y,z}[\vec e\,] / 2\pi^{2}$, such that
\be
Q[\vec e \,] = \frac{1}{2\pi^{2}} \sum_{\la wxyz \ra} V_{w,x,y,z}[\vec e\,]
\in \Z \ .
\ee
Remarkably, it was only in 2012 that a set of formulae was elaborated
which allow for the computation of $V_{w,x,y,z}[\vec e\,]$ \cite{Murakami}.
It can be numerically computed in this manner \cite{Nava}, but a more
efficient alternative is selecting some reference point on $S^{3}$
and counting how many spherical tetrahedra enclose it in an oriented
sense (we tested  extensively the equivalence of these two methods).

\subsection{Monte Carlo simulation}

As we anticipated in Section 1, the goal is the generation of numerous
configurations in accordance with the probability distribution
$p[\vec e\, ] = \frac{1}{Z} \exp (-S[\vec e \, ])$.

We start from an arbitrary initial configuration $[\vec e \, ]$ and generate
a long Markov chain
$[\vec e \, ] \to [\vec e \,'] \to [\vec e \, ''] \to \dots$
(each new configuration solely depends on the previous one, plus
some random numbers).
The conditions for the algorithm to be correct are {\em ergodicity}
(each configuration is accessible in a finite number of steps) and
{\em detailed balance}: the transition probabilities between two
configurations obey
\be
\frac{p[\vec e \to \vec e\,']}{p[\vec e\,' \to \vec e\,]} =
\frac{p[\vec e\,']}{p[\vec e\,]} = \exp ( S[\vec e\,]-S[\vec e\,']) \ .
\ee

One begins with the {\em thermalization:} first a large number of
configurations are skipped, until we reach thermal equilibrium
(and therefore independence of the initial configuration). Then
we perform numerical measurements on configurations, which have to
be sufficiently separated in the Markov chain to be statistically
independent from each other. To assure this property, we measure
the (exponential) auto-correlation ``time'' $\tau$; it is very similar
for the different observables involved (see below). We are on the
safe side with a measurement separation $\geq 2 \tau$.

$\tau$ grows rapidly next to the critical temperature. For typical
algorithms it diverges at $T_{\rm c}$ in infinite volume, and
the increase when $T$ approaches $T_{\rm c}$ is exponential:
this phenomenon is known as {\em critical slowing down}.

By definition, also the correlation length diverges at a critical
point, $\xi \to \infty$. Thus the spins are strongly correlated over
long distances (in lattice units), which explains that it becomes
hard to significantly modify a configuration (while respecting
detailed balance).

For the O($N$) models, the Wolff cluster algorithm \cite{Wolff}
is the most efficient, known simulation procedure. It does not
update single spins, but entire clusters of them are reflected at
some random hyper-plane in spin-space (they are ``flipped'').
The clusters are formed in a subtle manner, such that the algorithm
fulfills the aforementioned conditions
of ergodicity and detailed balance.

We used the multi-cluster version (but we also checked its consistency
with the single-cluster algorithm). One multi-cluster update step means
that the entire configuration is divided into clusters, which are flipped
with the appropriate probability. The availability of this highly
efficient algorithm is another benefit of the O(4) model as an
effective theory; no efficient cluster algorithm is known in gauge
theory. We take the chemical potential $\mu_{B}$ into account by
adjusting the cluster flip probability, along the lines of Ref.\
\cite{ForKas}. This method works consistently, but when $\mu_{B,{\rm lat}}$
increases, the peak height of $\tau$ grows rapidly.

\begin{figure}[H]
\centering
\vspace*{-2mm}
\includegraphics[scale=0.5]{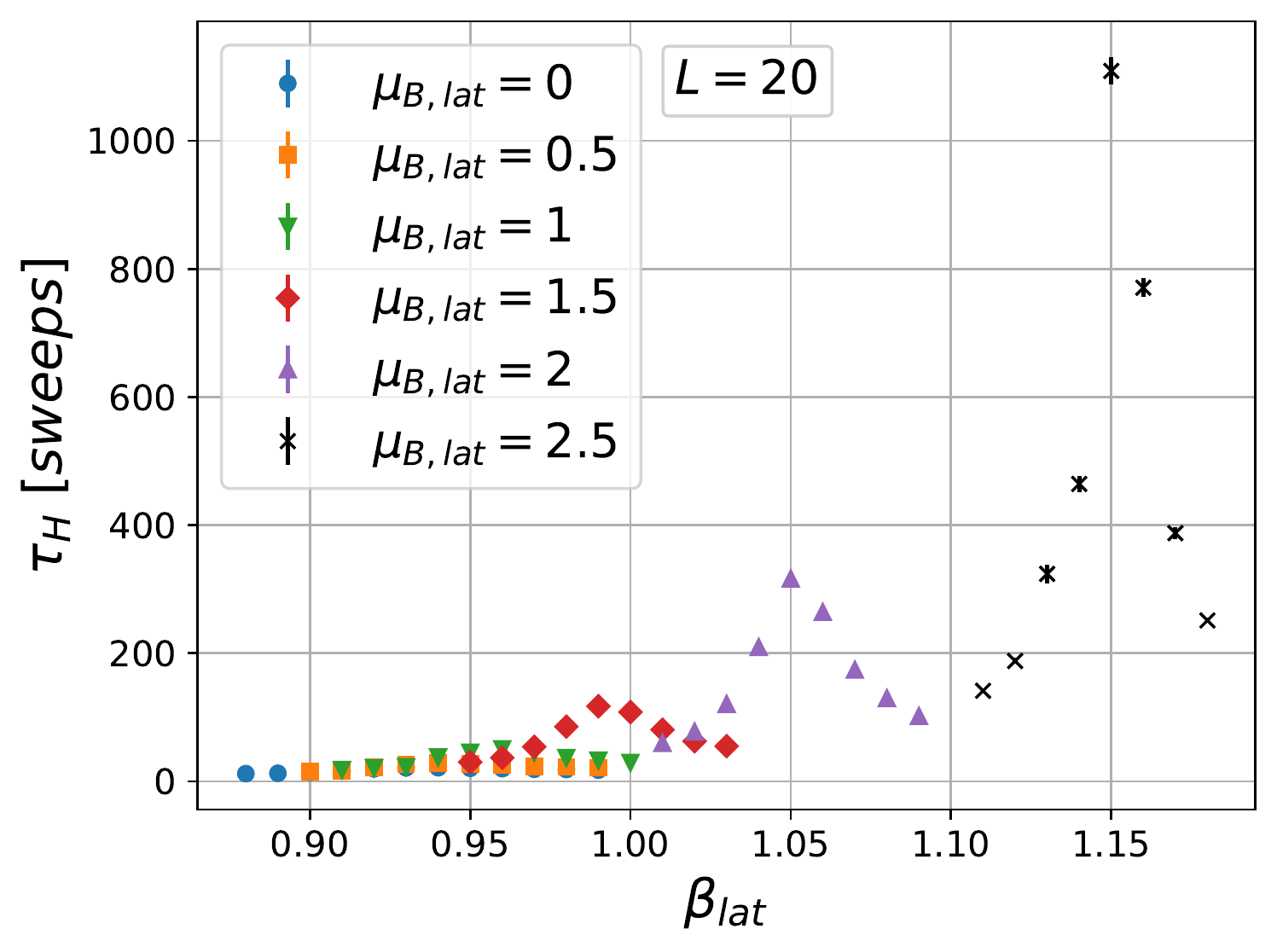}
\includegraphics[scale=0.5]{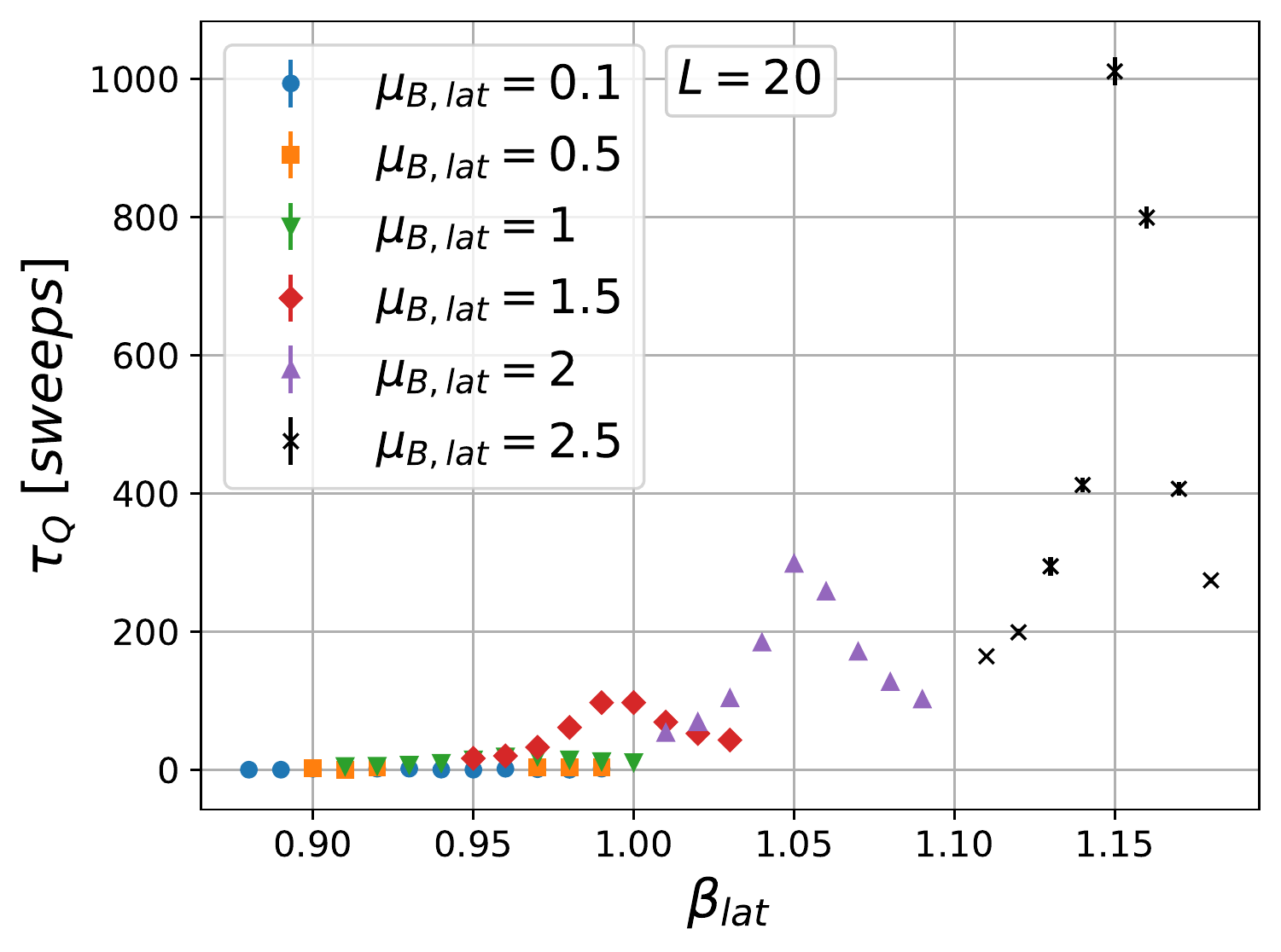}
\vspace*{-2mm}
\caption{The auto-correlation ``time'' with respect to the energy,
  $\tau_{H}$ (top) and with respect to the topological charge,
  $\tau_{Q}$ (bottom), expressed in units of multi-cluster update
  steps (``sweeps''). These values are measured in the chiral limit
  ($\vec h_{\rm lat} =\vec 0$) by the exponential decay of the
  auto-correlation.}
\vspace*{-3mm}
\label{taufig}
\end{figure}

This is illustrated in Fig.\ \ref{taufig} for the auto-correlation
``times'' with respect to the energy, $\tau_{H}$, and the
topological charge, $\tau_{Q}$. They are very similar, thanks to the
cluster algorithm (for single-spin update algorithms, $\tau_{Q}$ tends to
be much larger and to restrict the feasibility of conclusive simulations).

We see that not even the cluster algorithm completely overcomes the problem
of critical slowing down. This difficulty has limited our numerical
study so far to $\mu_{B,{\rm lat}} \leq 2.5$.
On the other hand, the sharp peaks of $\tau$ provide a first
estimate of the critical value $\beta_{\rm c,lat}$.

\section{Results for the phase diagram in the chiral limit}

We begin with the case $\vec h_{\rm lat} = \vec 0$, which corresponds to zero
quark and pion masses. Before showing our simulation results,
which are based on Ref.\ \cite{Edgar}, we address
the conversion from lattice units to physical units. This requires
some reference quantity as an input. Here, we refer to the critical
temperature $T_{\rm c} = 1/\beta_{\rm c}$ at $\mu_{B} = 0$.
In the 3d O(4) model on the lattice, it was measured to high precision
\cite{Kanaya,Oevers,Bielefeld}; we are going to refer to
$\beta_{\rm c,lat} = 0.9359(1)$. We match this result to
$T_{\rm c} \simeq 132~{\rm MeV}$, the value obtained in chiral
lattice QCD \cite{Ding19} (cf.\ Section 1), which suggests
\be
\mu_{B} = \frac{\beta_{\rm c,lat}}{\beta_{\rm c}} \ \mu_{B,{\rm lat}}
\approx 124~{\rm MeV} \ \mu_{B,{\rm lat}} \ .
\ee
Our simulation parameters are
$$
\mu_{B,{\rm lat}} = 0,\ 0.1,\ 0.2\ \dots 1.5; \ 2,\ 2.5
\ \Leftrightarrow \ \mu_{B} = 0 \dots 309~{\rm MeV} \ .
$$
The lattice volumes are cubic, $V = L^{3}$, $L=10,\ 12,\ 16,\ 20$,
and, if not indicated otherwise, we will show results for $L=20$.
The $\beta_{\rm lat}$-values are chosen such that $\beta_{\rm c,lat}$ can
be identified --- this had to be explored at each $\mu_{B,{\rm lat}}$.

We measured observables which are given by first and second derivatives
of the free energy $F= -T \ln Z$. According to Ehrenfest's scheme, a
discontinuity in the $n^{\rm th}$ derivative (in the large-$L$ limit)
characterizes an  $n^{\rm th}$ order phase transition. We monitor the
critical line and search in particular for a possible CEP, as motivated
in Section 1. Each measurement is based on $10^{4}$
(thermalized and decorrelated) configurations.

\begin{figure}[H]
\centering
\includegraphics[scale=0.5]{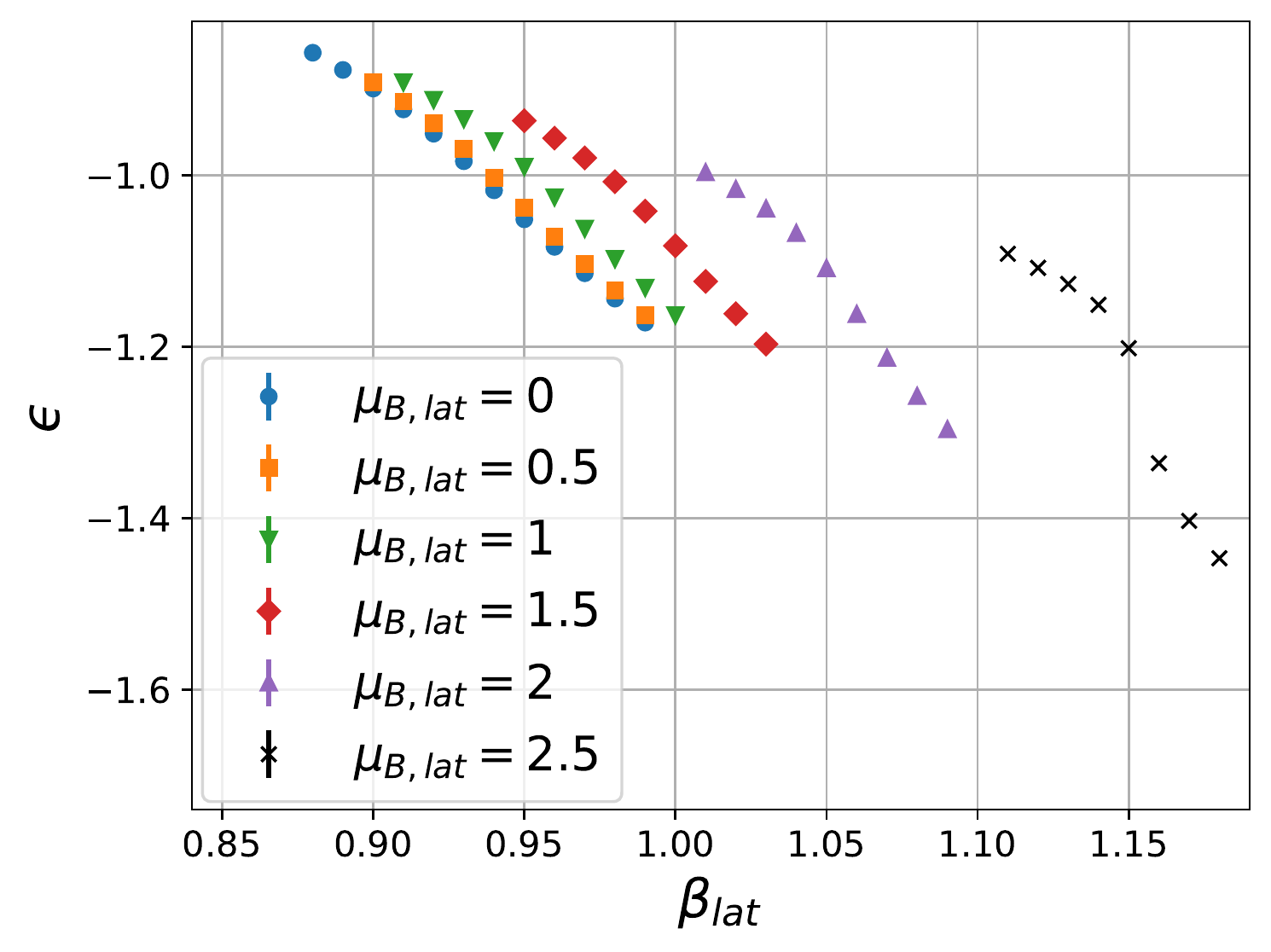}
\includegraphics[scale=0.5]{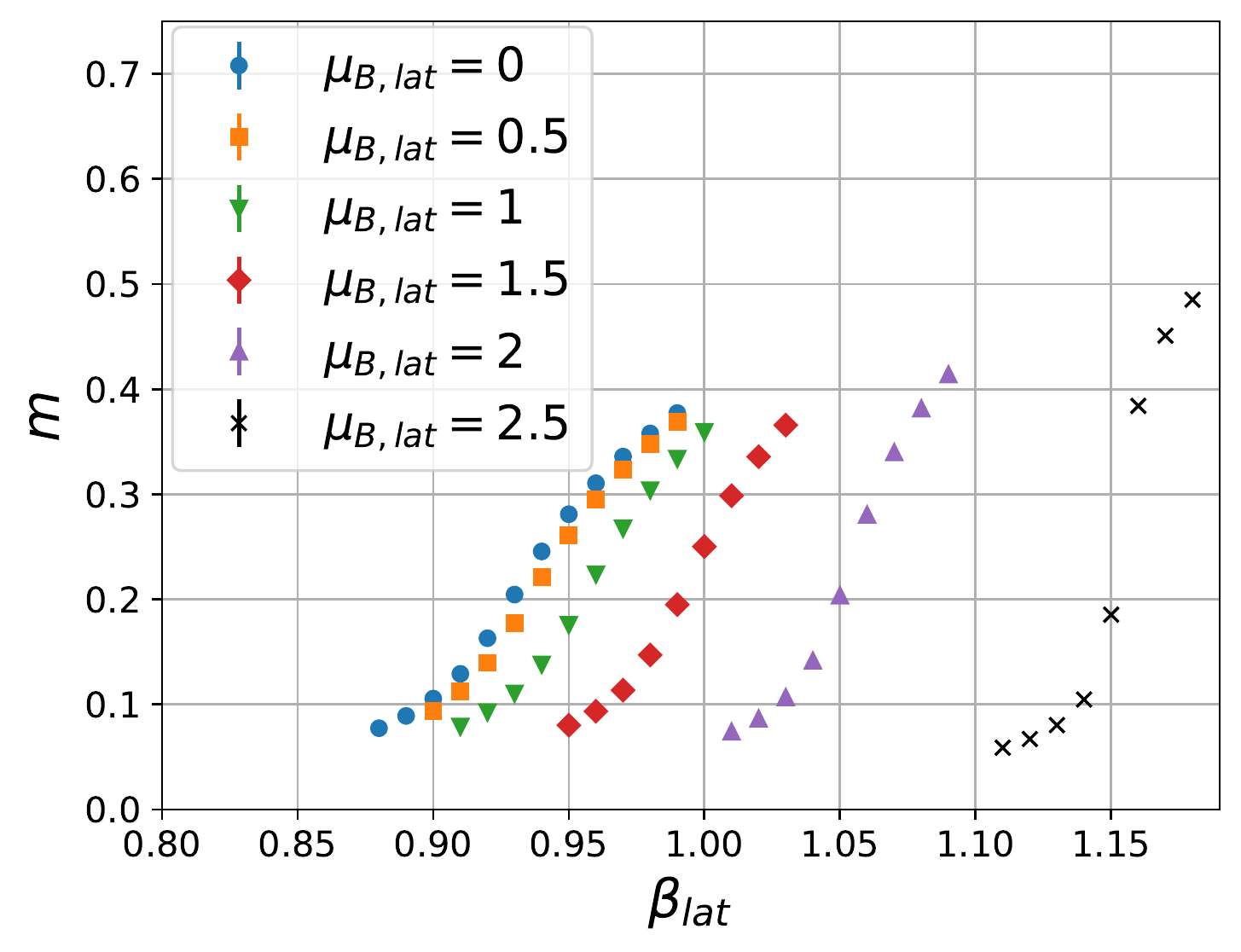}
\includegraphics[scale=0.5]{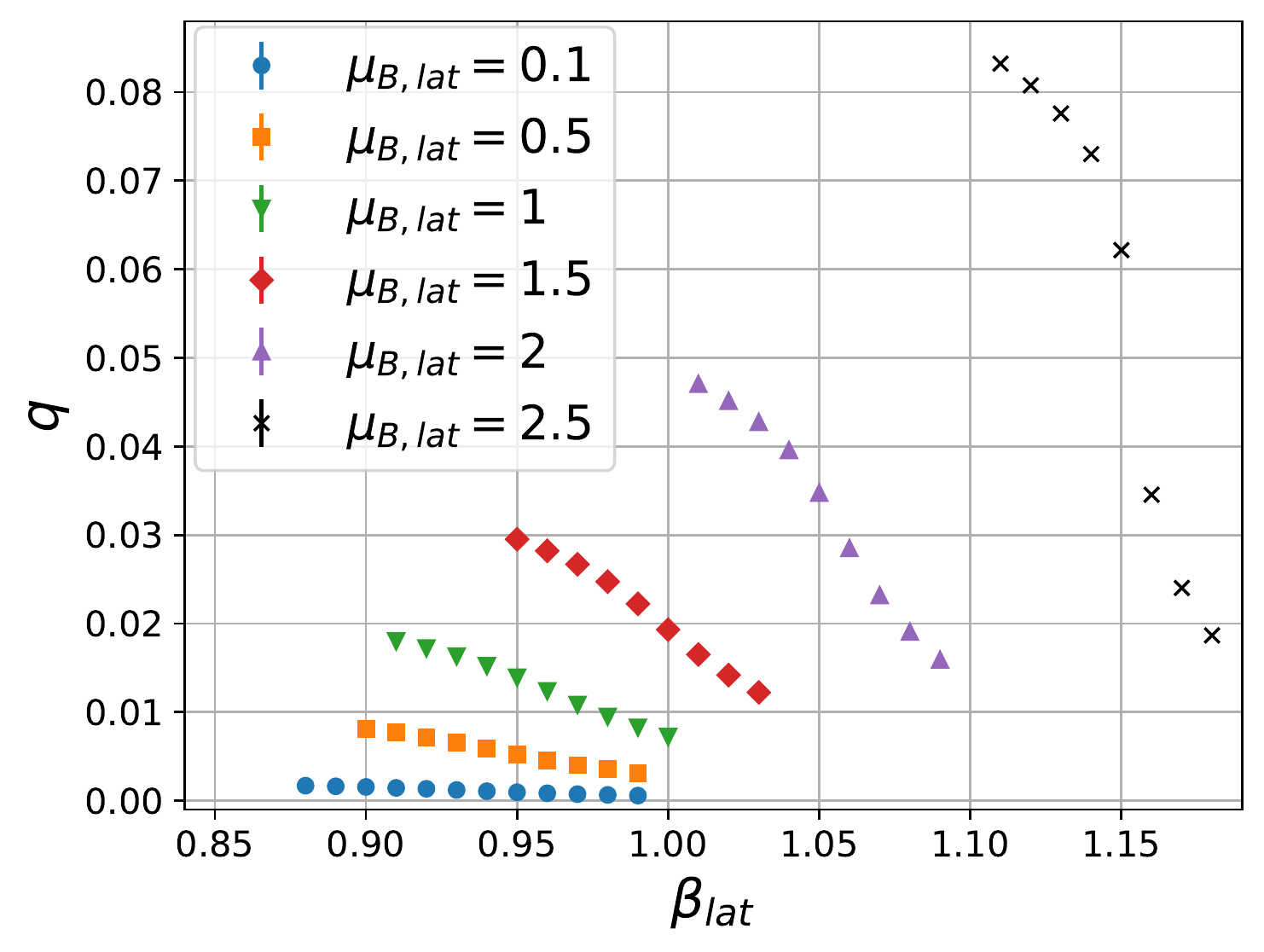}
\caption{Energy density $\epsilon$, magnetization density $m$
  and topological density $q$,
  at $L=20$, $\vec h=\vec 0$ and $\mu_{B,{\rm lat}} = 0 \dots 2.5$.}
\vspace*{-0.2cm}
\label{epsmdensity}
\end{figure}
Figure \ref{epsmdensity} shows the energy density $\epsilon$,
the magnetization density $m$ (the order parameter) and the
topological density $q$, which are all given by first derivatives
of $F$,\footnote{In eqs.\ (\ref{epsmqdef}), (\ref{cVdef}) and
  (\ref{chimdef}), $\beta$ and $\mu_{B}$ are understood as
  $\beta_{\rm lat}$ and $\mu_{B,{\rm lat}}$ for the interpretation of
  our data.}
\bea
\epsilon &=& \la H \ra / V = \frac{1}{V} \partial_{\beta} (\beta F) \ , \nn \\
m &=& \la |\vec M | \ra /V \ ,
\ \la \vec M  \ra = \Big\la \sum_{x} \vec e_{x} \Big\ra
= - \partial_{\vec h} F \ , \nn \\
q &=& \la Q \ra /V \ , \quad \la Q \ra = - \partial_{\mu_{B}} F \ .
\label{epsmqdef}
\eea
Increasing $\mu_{B,{\rm lat}}$ favors more topological windings.
This enhances $q$ and also $\epsilon$, but it reduces $m$, since
the configurations are further away from a uniform structure.
Clearly, increasing $\beta_{\rm lat}$ has the opposite effect. 
In all three plots we see
intervals of maximal slope, which move to large $\beta_{\rm lat}$
when $\mu_{B,{\rm lat}}$ grows: this indicates the approximate
value of $\beta_{\rm c,lat}$, in agreement with the peaks of
the auto-correlation ``times'' in Fig.\ \ref{taufig}.
\begin{figure}[H]
\centering
\includegraphics[scale=0.5]{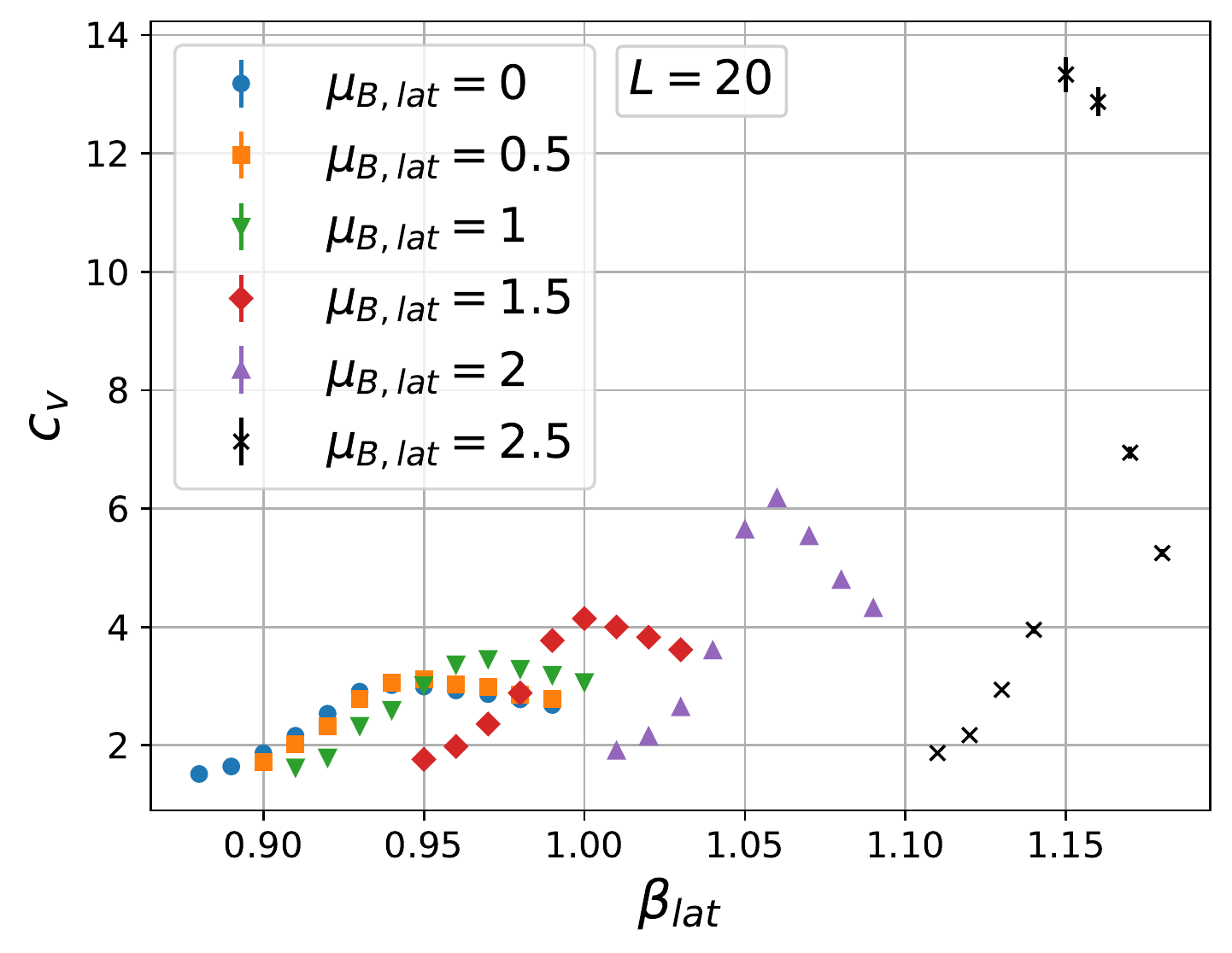}
\includegraphics[scale=0.5]{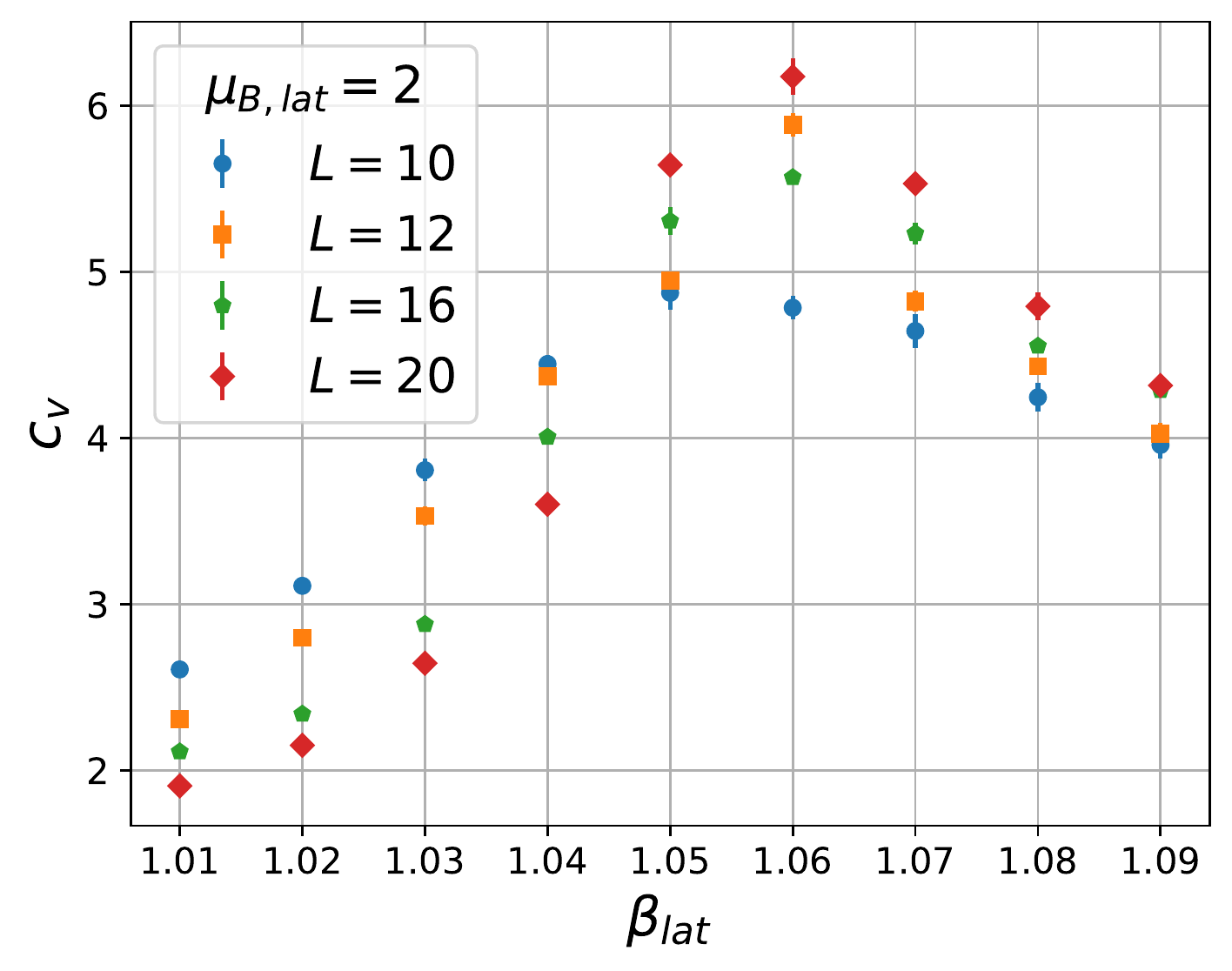}
\caption{The specific heat $c_{V}$ at $L=20$ (top) and at $\mu_{B,{\rm lat}}=2$
  (bottom). The peaks hint at second order phase transitions in the
  large-$L$ limit. The $L$-dependence of their heights provides
information about the critical exponents $\alpha$ and $\nu$.}
\vspace*{-0.2cm}
\label{cVplot}
\end{figure}

For $\mu_{B,{\rm lat}} = 2.5$ these slopes are so strong that one
could even be tempted to interpret them as quasi-discontinuous
jumps, {\it i.e.}\ they could indicate discontinuous jumps in
the large-$L$ limit. That would be characteristic of a {\em first}
order phase transition, so at this point we wonder whether the
CEP has been attained already. This has to be clarified by the study
of further observables, which are given by second derivatives of
the free energy $F$.

In this respect, we first consider the specific heat $c_{V}$,
\be  \label{cVdef}
c_{V} = \frac{\beta^{2}}{V} ( \la H^{2} \ra - \la H \ra^{2} ) =
- \frac{\beta^{2}}{V} \partial^{2}_{\beta} (\beta F) \ .
\ee
In infinite volume it diverges at a second order phase transition.
In Fig.\ \ref{cVplot} (top) we see peaks with increasing height
when $\mu_{B,{\rm lat}}$ rises. This indicates that the phase transition
at these parameters, in infinite volume, is still second order.

This is more explicit in Fig.\ \ref{cVplot} (bottom), which compares
$c_{V}(\beta_{\rm lat})$ in different volumes.
The peak centers hardly depend on $L$, which makes their large-$L$
extrapolation simple.
Based on finite-size scaling one expects (assuming $L=\infty$ to
be a critical point) a peak height proportional to $L^{\alpha/\nu}$.
At $\mu_{B,{\rm lat}}=2$ we obtain for the ratio of these critical
exponents $\alpha / \nu \approx 0.2$. If we further assume
Josephson's scaling law $\alpha = 2 - d\nu$, we arrive at
$\alpha \approx 1/8$, $\nu \approx 5/8$. As a benchmark,
Ref.\ \cite{Kanaya} obtained at $\mu_{B,{\rm lat}}=0$ the value
$\nu \simeq 0.7479(90)$, which is in reasonable proximity.

\begin{figure}[H]
\centering
\includegraphics[scale=0.5]{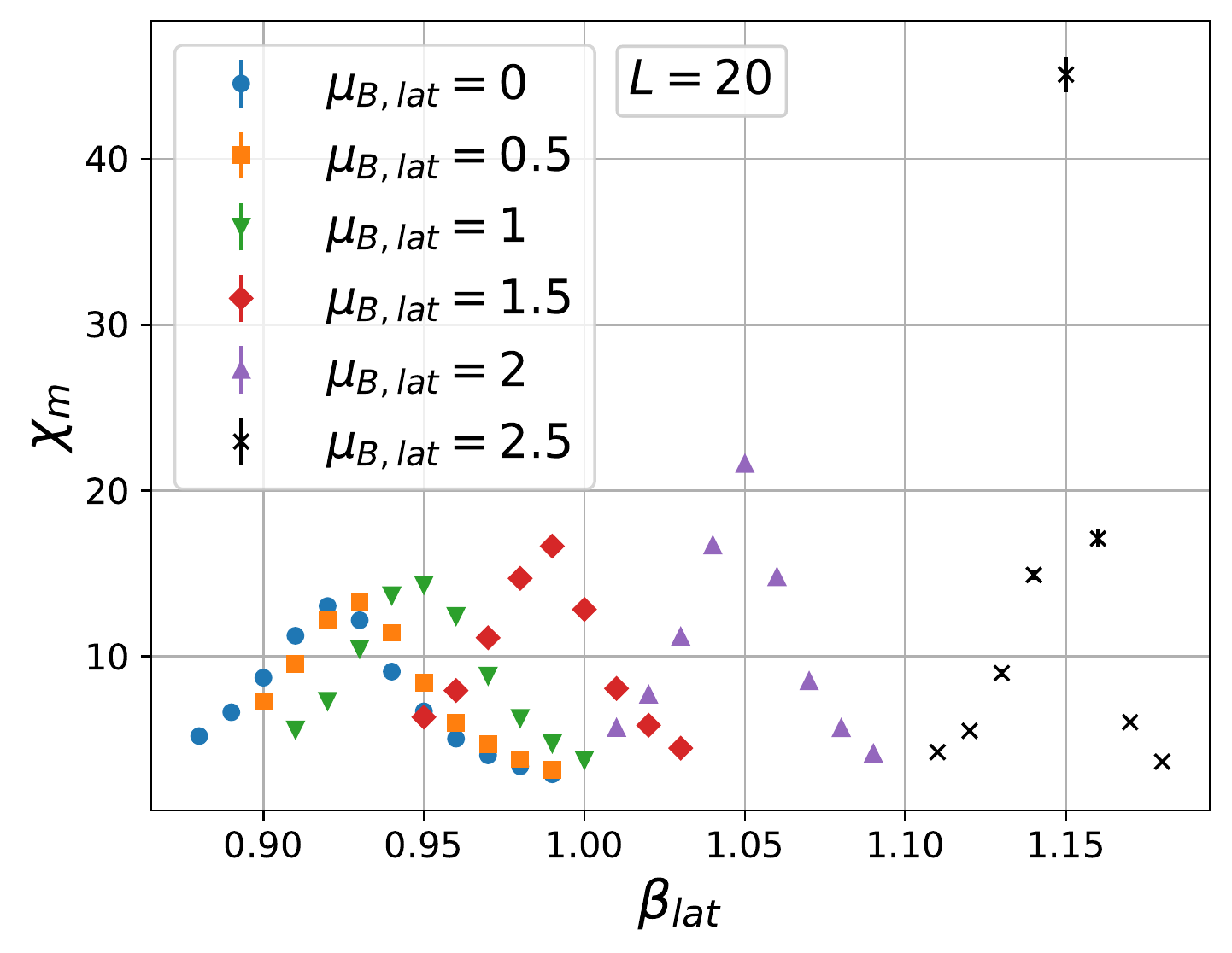}
\includegraphics[scale=0.5]{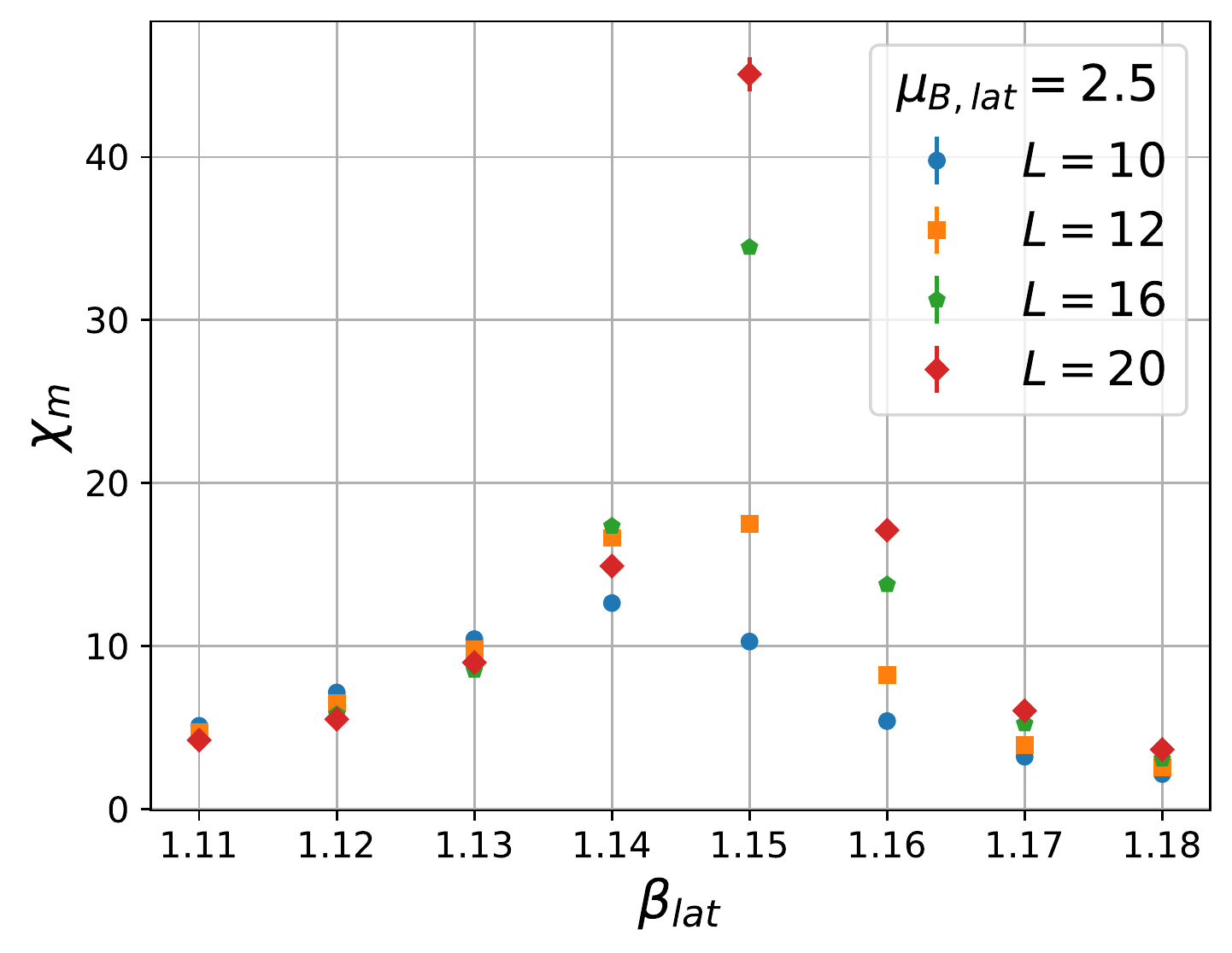}
\caption{The magnetic susceptibility $\chi_{m}$ at $L=20$ (top)
  and at $\mu_{B,{\rm lat}}=2.5$ (bottom). The peaks again hint at second
  order phase transitions in the large-$L$ limit. The $L$-dependence of
  their heights provides information about the ratio of
  critical exponents $\gamma/\nu$.}
\vspace*{-0.2cm}
\label{chimplot}
\end{figure}

Similarly, in Fig.\ \ref{chimplot} (top) we show results for the magnetic
susceptibility $\chi_{m}$ at $L=20$,
\be  \label{chimdef}
\chi_{m} = \frac{\beta}{V} \left(
\la \vec M^{2} \ra - \la |\vec M| \ra^{2} \right) \sim
- \frac{\beta}{V} \partial^{2}_{\vec h} F \ .
\ee
(In a numerical study, the subtracted term is only sensible with $|\vec M|$,
see {\it e.g.}\ Ref.\ \cite{Binder}. The right-hand side, however, is the
standard formula, without absolute value.) It also diverges at $T_{\rm c}$ in
infinite volume, hence its peaks in finite volume are another indicator
of a second order phase transition. They are strongest at
$\mu_{B,{\rm lat}} \geq 1$, which supports the scenario that we are still
following a critical line.

The plot in Fig.\ \ref{chimplot} (bottom) shows results for $L=10 \dots 20$
at $\mu_{B,{\rm lat}} = 2.5$. Here, the peak temperature visually moves with
$L$, and the large-$L$ extrapolation is compatible with the estimates
for $T_{\rm c}$ based on the previous criteria. In this case, one
expects ${\rm (peak~height)} \propto L^{\gamma / \nu}$. In the range
$\mu_{B,{\rm lat}} \leq 1.5$ this yields $\gamma / \nu = 1.9(2)$
\cite{Edgar}, in agreement with the precise value
$\gamma / \nu = 1.970$ at $\mu_{B,{\rm lat}} =0$ \cite{Bielefeld}.

\begin{figure}[H]
\vspace*{-2mm}
\centering
\includegraphics[scale=0.5]{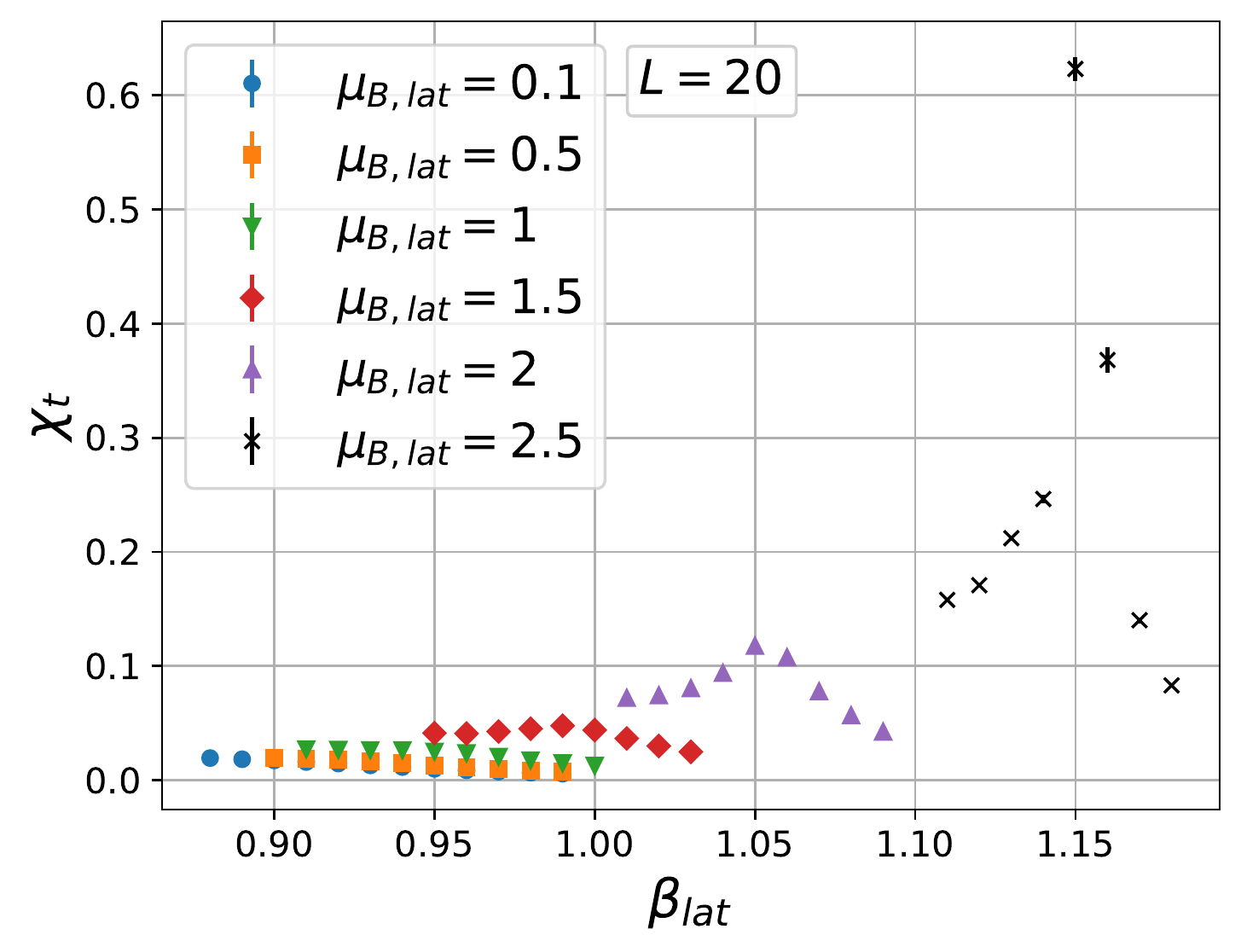}
\caption{The topological susceptibility $\chi_{\rm t}$ at $L=20$.
  The peaks at $\mu_{B,{\rm lat}} = 2$ and $2.5$ further support the scenario
  of a second order phase transition.}
\vspace*{-0.2cm}
\label{chitplot}
\end{figure}

Figure \ref{chitplot} adds results about the topological susceptibility
\be
\chi_{\rm t} = \frac{1}{V} \left( \la Q^{2} \ra - \la Q \ra^{2} \right)
= - \frac{1}{V} \partial_{\mu_{B}}^{2} F \ .
\ee
Again we observe peaks (they are obvious at
$\mu_{B,{\rm lat}} \geq 2$), at temperatures which are consistent
with the previous determinations of $T_{\rm c}$. 
Regarding the peak height, in analogy to Figs.\ \ref{cVplot} and
\ref{chimplot}, one might define a critical exponent $\zeta$ by the
relation $\chi_{\rm t}(T_{\rm c}) \propto L^{\zeta/\nu}$, for which we obtain
{\it e.g.}\
\be
\frac{\zeta}{\nu} \approx \left\{ \begin{array}{cc}
  0.2 & \mu_{B,{\rm lat}=0} \\
  0.3 & \mu_{B,{\rm lat}=1} \end{array} \right. \ .
\ee

\begin{figure}[H]
\centering
\includegraphics[scale=0.57]{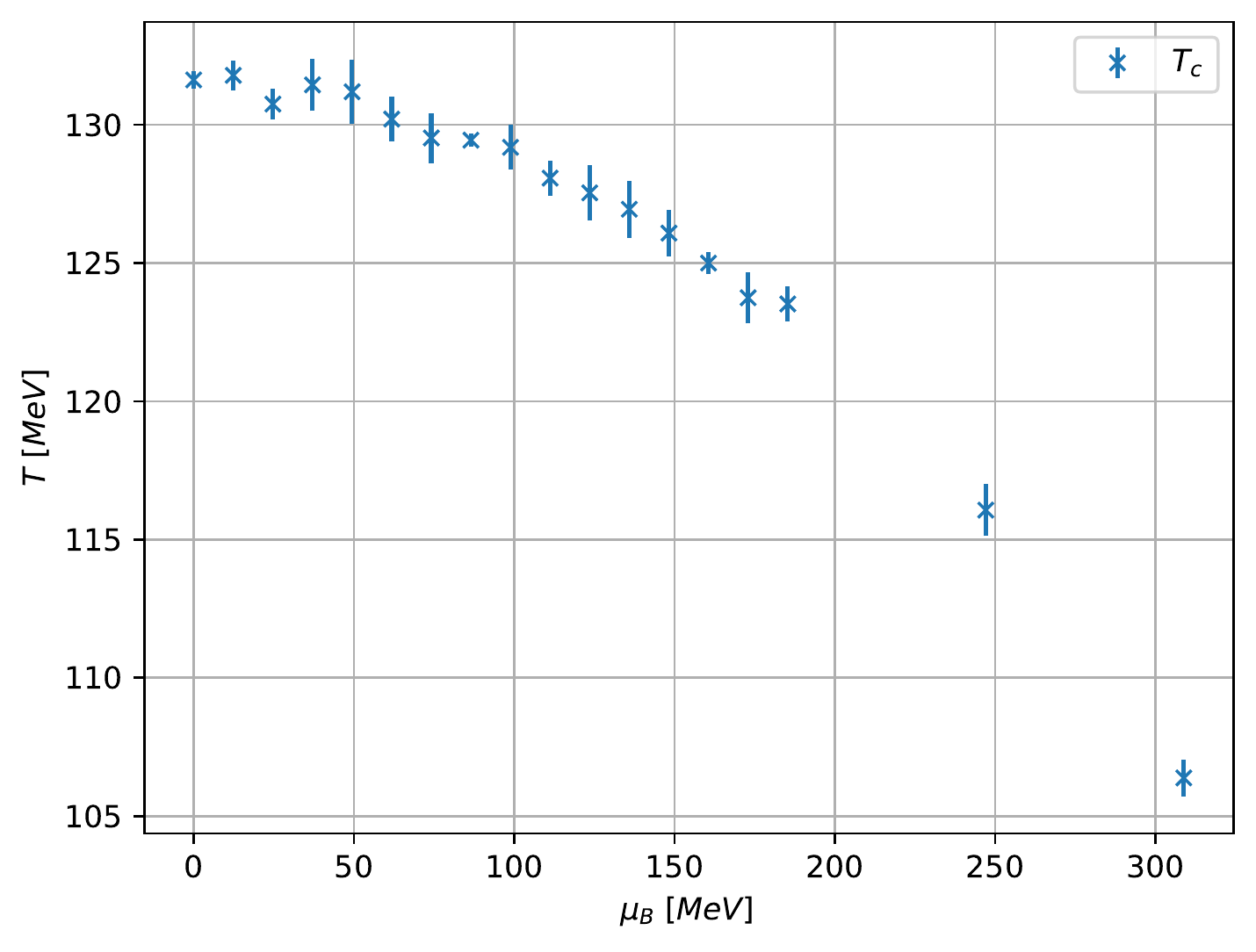}
\caption{Conjectured phase diagram of 2-flavor QCD in the chiral limit.}
\vspace*{-0.2cm}
\label{phasediachiral}
\end{figure}

Taking all these results for quantities given by second derivatives
of $F$ together, strongly supports the scenario of a second order
phase transition, all the way up to $\mu_{B,{\rm lat}} = 2.5$.
If we combine all the indications for the values of
$\beta_{\rm c,lat}(\mu_{B,{\rm lat}})$ (peaks and steepest slopes),
extrapolate to the thermodynamic limit $L \to \infty$, and convert
the outcome into physical units, we arrive at our conjecture for the
chiral phase diagram in Fig.\ \ref{phasediachiral}.
We see that $T_{\rm c}$ decreases monotonically with increasing
baryon density, as generally expected.
According to this diagram, a possible CEP should be
located at $\mu_{B} > 309~{\rm MeV}$ and $T < 106~{\rm MeV}$.

\section{Results with light quarks}

We repeat that the O(4) model represents an effective theory for
2-flavor QCD, where a ``magnetic field'' $h = |\vec h|$ plays a
role analogous to a degenerate quark mass $m_{q} =m_{u}=m_{d}$.
This is the parameter which adds some explicit symmetry breaking,
and gives mass to the pions (this is well-known in Chiral Perturbation
Theory).

We simulated at two values of this parameter, in lattice units they
amount to $h_{\rm lat} = 0.14$ and $h_{\rm lat} = 0.367$.
For the conversion between lattice units and physical units, we now
refer to the phenomenological, pseudo-critical crossover temperature
$T_{\rm pc} \simeq 155~{\rm MeV}$ at zero baryon density
\cite{BazavovBhattacharya}.
Our simulation results for $T_{\rm pc,lat}$ are ambiguous, as expected for
a crossover, see below. We anticipate the mean values in lattice units
at $\mu_{B,{\rm lat}}=0$ (at this point without uncertainties),
\bea
\bar T_{\rm pc,lat} &=& \left\{ \begin{array}{cc}
  1.172 & h_{\rm lat} = 0.14 \\
  1.273 &  ~\, h_{\rm lat} = 0.367
\end{array} \right. \nn \\
\Rightarrow \ \mu_{B} &=& \frac{T_{\rm pc}}{\bar T_{\rm pc,lat}} \ \mu_{B,{\rm lat}}
\nn \\
&=& \left\{ \begin{array}{cc}
  132~{\rm MeV} \ \mu_{B,{\rm lat}} & h_{\rm lat} = 0.14 \\
  122~{\rm MeV} \ \mu_{B,{\rm lat}} & ~\, h_{\rm lat} = 0.367
\end{array} \right. . \qquad
\eea
Still following the analogy to QCD, we interpret the chiral
symmetry breaking parameter as
\be
h = \frac{T_{\rm pc}^{4}}{\bar T_{\rm pc,lat}^{4}} \ h_{\rm lat} = m_{q} \Sigma \ ,
\ee
with the chiral condensate $\Sigma = -\la \bar \psi \psi \ra
\approx (250~{\rm MeV})^{3}$, which allows us to estimate
the physical values of the quark mass,
\be
m_{q} \approx  \left\{ \begin{array}{cc}
  3~{\rm MeV} &  h_{\rm lat} = 0.14 \\
  5~{\rm MeV} & ~\, h_{\rm lat} = 0.367
\end{array} \right. \ .
\ee

\begin{figure}[H]
\centering
\includegraphics[scale=0.7]{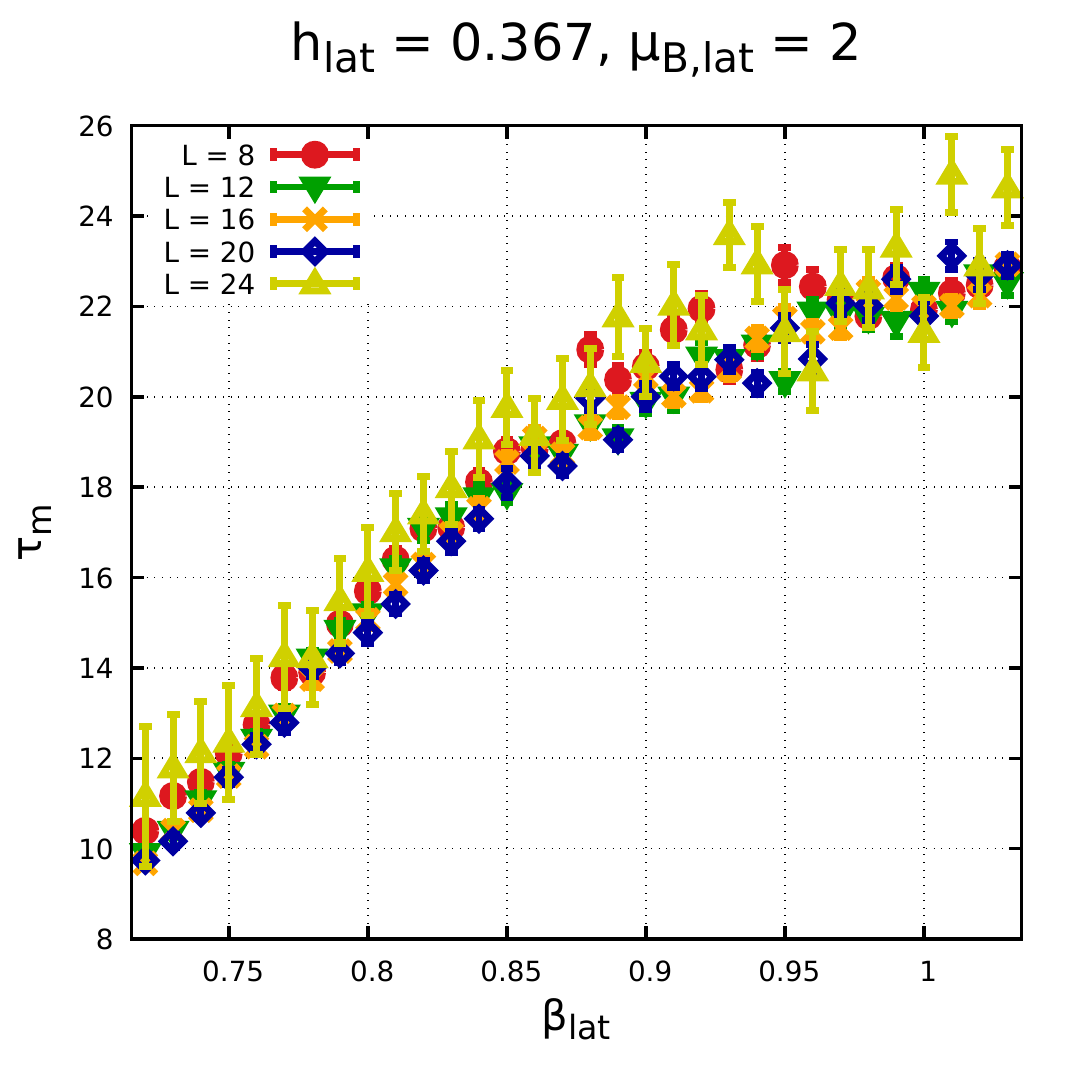}
\caption{Auto-correlation ``time'' with respect to the magnetization
in units of multi-cluster update steps. There is only a minor dependence
on the size $L$, and no critical slowing down.}
\vspace*{-0.2cm}
\label{tauplotm}
\end{figure}

We include also this symmetry-breaking term in the cluster algorithm by
modifying the cluster-flip probability, as described in Refs.\
\cite{Wang,JAGH}. The magnitude of the auto-correlation time $\tau$
is strongly alleviated compared to Section 3, see Fig.\ \ref{tauplotm},
since the critical line is replaced by a crossover. $\tau$ does not
diverge at $\beta_{\rm pc}$ in infinite volume, hence there is no critical
slowing down in this case. Thus the massive model is computationally
less demanding, which allowed us to include $L=24$, and larger volumes
are accessible as well; this is work in progress.

So far our simulation parameters are
\be
V = L^{3} \ , \ L=8,\ 12,\ 16,\ 20,\ 24 \ ; \quad
\mu_{B,{\rm lat}} = 0 \dots 2 \ .
\ee
At $h_{\rm lat}=0.14$ we have additional data at $\mu_{B,{\rm lat}} = 2.5$,
which corresponds to $\approx 330~{\rm MeV}$.

In the following,
we are going to show results for $h_{\rm lat}=0.367$ at $\beta_{\rm lat}$-values
in the crossover region, again based on $10^{4}$ measurements at each
parameter set, for similar observables as in Section 3,
following Ref.\ \cite{JAGH}. The results at $h_{\rm lat}=0.14$ look alike;
they will be included in our final conjecture about the phase diagram
in the massive case.

Figure \ref{tauplotm} shows the auto-correlation ``time'' with respect to
the magnetization, $\tau_{m}$: the absence of critical slowing down is
obvious, so we are on the safe side if we separate the measurements by
45 multi-cluster update steps.
On the other hand, in contrast to the chiral case, $\tau_{m}$
does not provide a first estimate for $\beta_{\rm pc,lat}$.

\begin{figure}[H]
  \centering
  \includegraphics[scale=0.7]{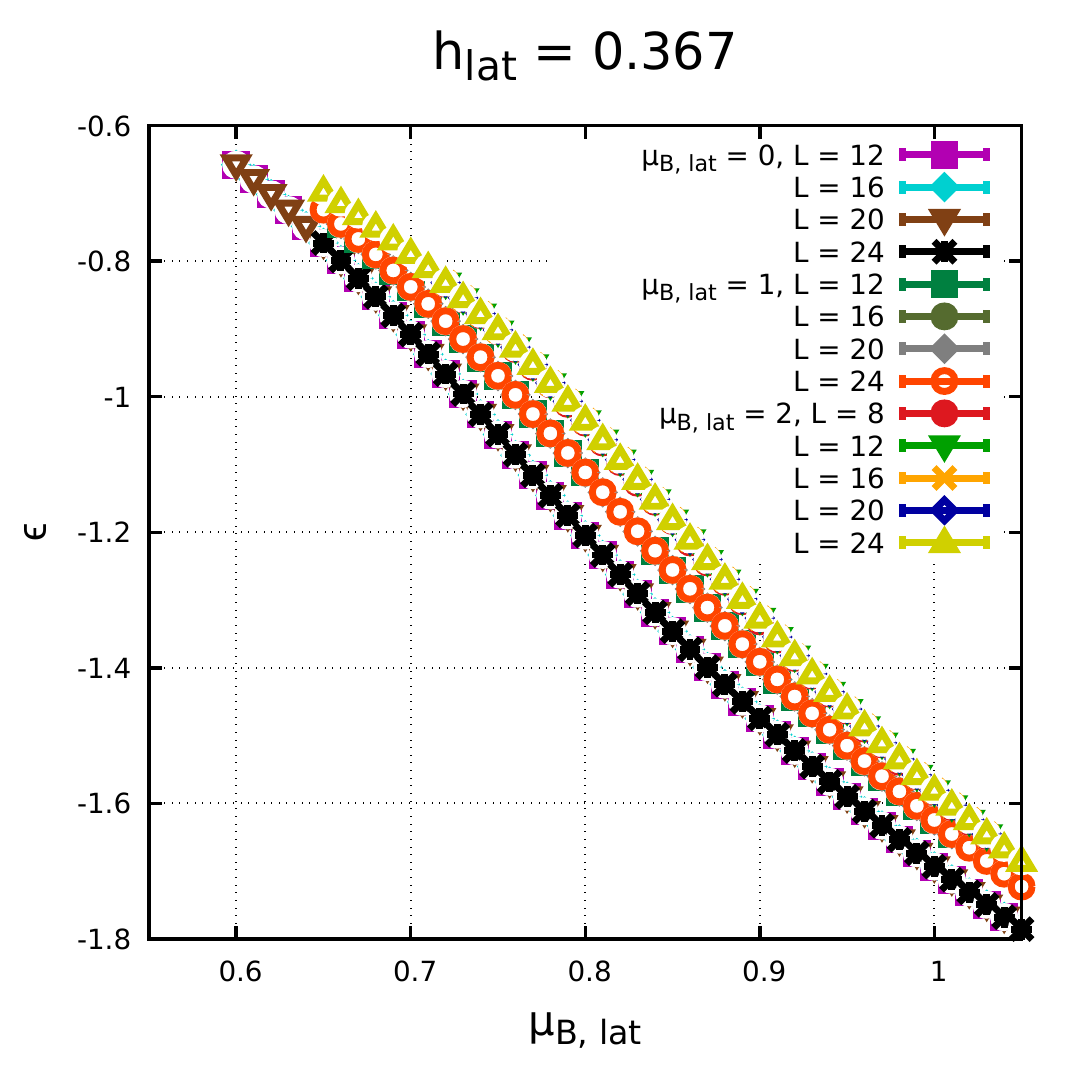}
\includegraphics[scale=0.7]{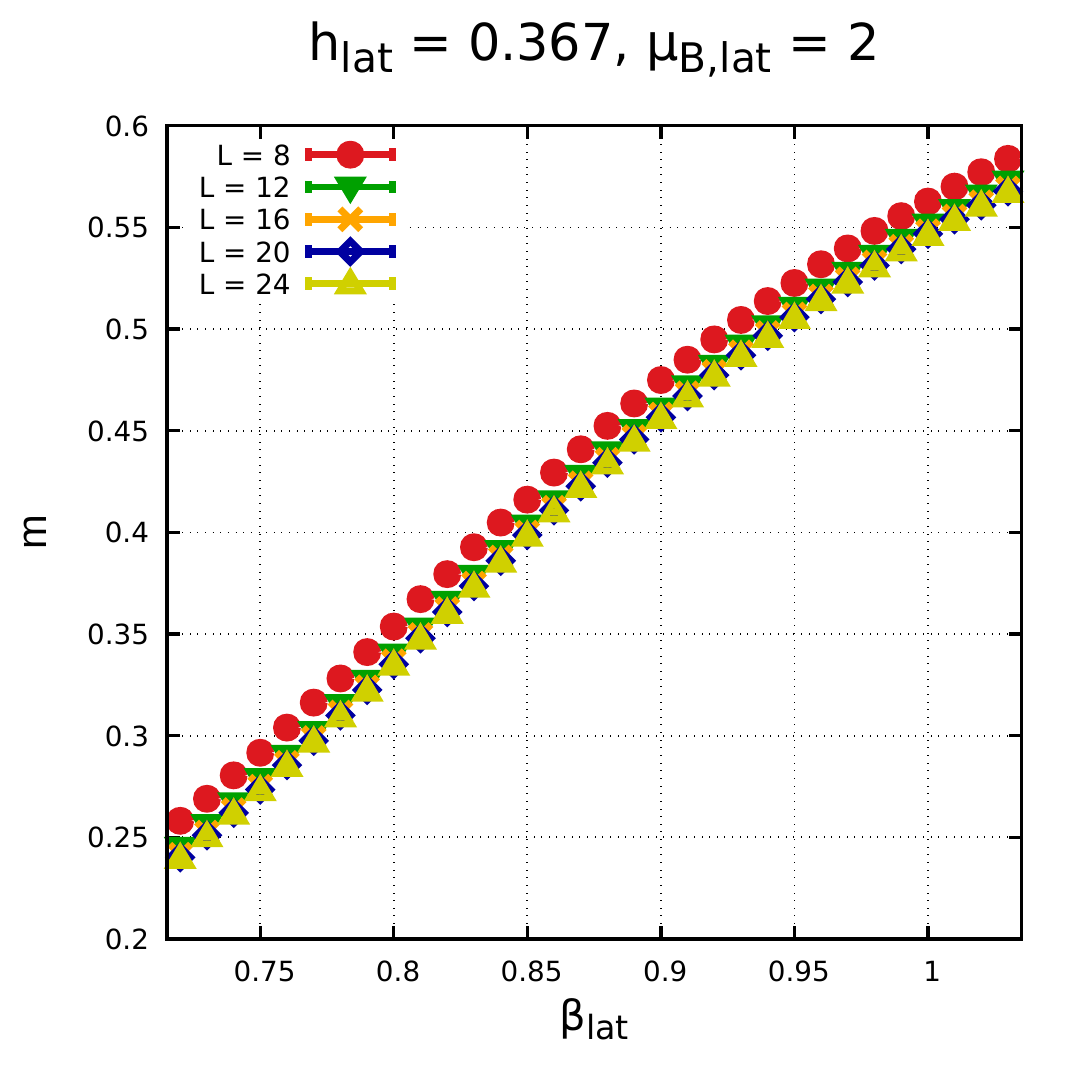}
\vspace*{-1mm}
\caption{Energy density $\epsilon$ and magnetic density $m$
  at $h_{\rm lat}=0.367$. We see shifts depending on $\mu_{B,{\rm lat}}$,
  but no interval of extraordinary slopes, since the second order
  phase transition (at $h_{\rm lat}=0$) is now washed out to a crossover.
  Modest finite-size effects are visible for $m$.}
\vspace*{-0.2cm}
\label{epsmdensm}
\end{figure}
In Fig.\ \ref{epsmdensm} we proceed to the energy density
$\epsilon$ and the magnetization density $m$, cf.\ eqs.\
(\ref{epsmqdef}). Only for $m$ modest finite-size effects
are visible, but changing $\mu_{B,{\rm lat}}$ causes a shift in $\epsilon$.
In either case, there is no interval of an extraordinary slope
(which would increase with $L$); this confirms that we are not
dealing with a phase transition.

\begin{figure}[H]
\centering
\vspace*{-2mm}
\includegraphics[scale=0.7]{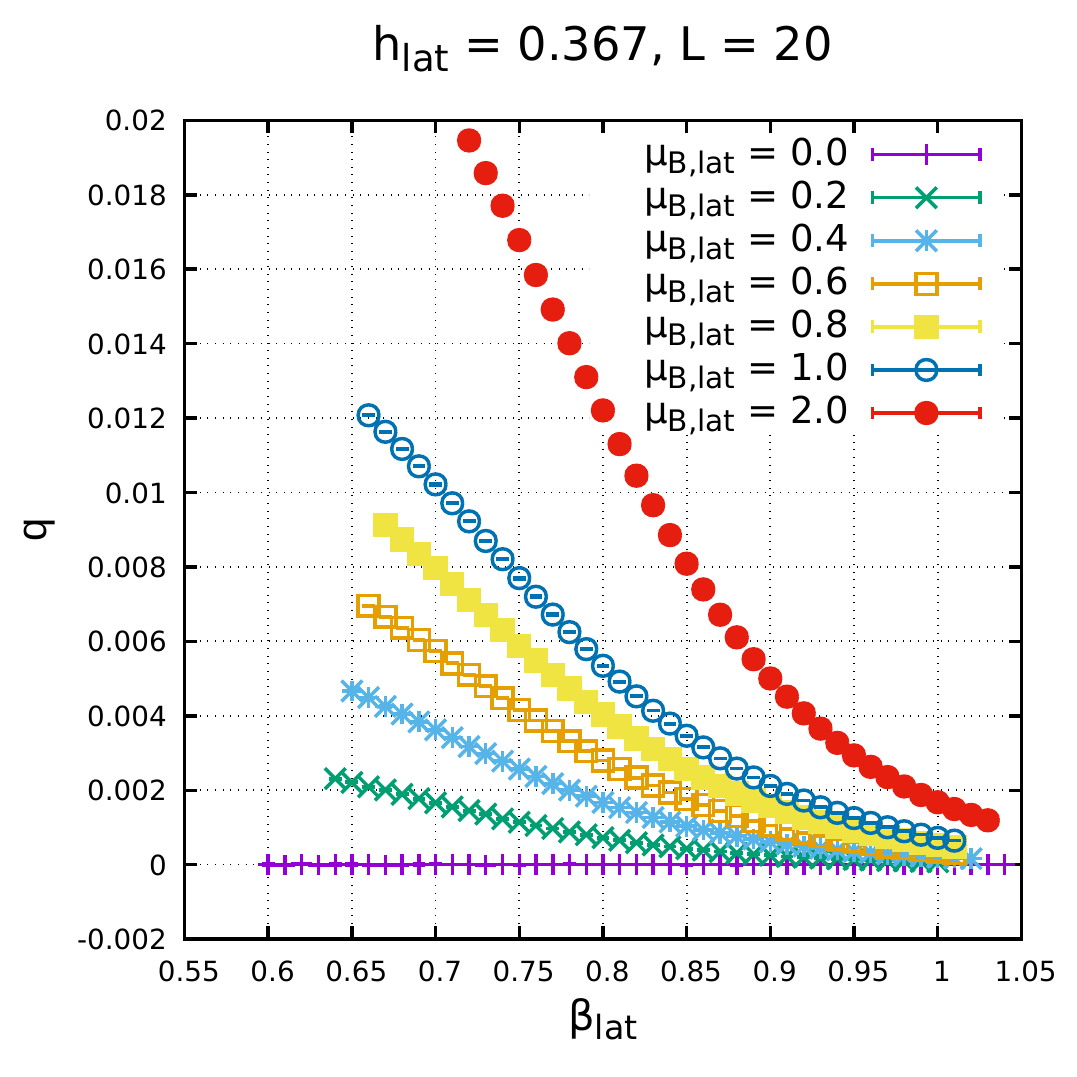}
\includegraphics[scale=0.7]{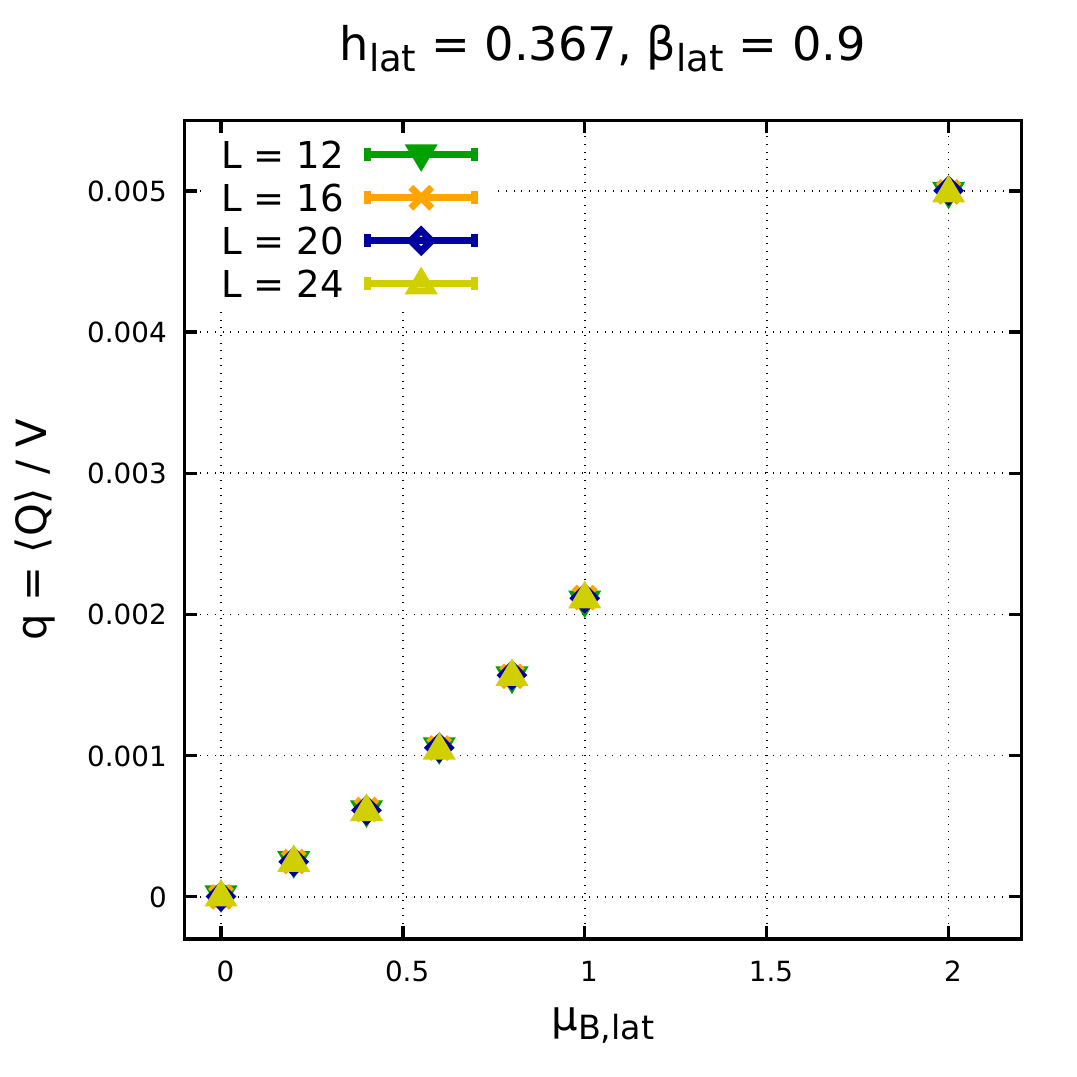}
\vspace*{-1mm}
\caption{The topological density $q = \la Q \ra /V$ as function
  of $\beta_{\rm lat}$ at $L=20$ (top), and at $\beta_{\rm lat} = 0.9$
  in different volumes (bottom). We see how an increasing $\mu_{B,{\rm lat}}$
  enhances $q$, whereas increasing $\beta_{\rm lat}$ suppresses it.
  At $L\geq 12$ there are hardly any finite-size effects on $q$.}
\vspace*{-0.2cm}
\label{qdensm}
\end{figure}

The topological density $q = \la Q \ra /V$ is illustrated in Fig.\
\ref{qdensm} in two ways, as a function of $\beta_{\rm lat}$ and of
$\mu_{B,{\rm lat}}$. At $\mu_{B,{\rm lat}}=0$, parity symmetry implies $q=0$.
Obviously, $\mu_{B,{\rm lat}}>0$ enhances $q$, while increasing $\beta$
suppresses topological windings, and for $L\geq 12$ it is hardly
affected by finite-size effects.

We add that we did not observe any maxima in the topological
susceptibility $\chi_{\rm t}$ in the interval if $\beta_{\rm lat}$
that we explored. So in the massive case, $\chi_{\rm t}$ is not
helpful in view of the phase diagram, hence we do not include its plot.

\begin{figure}[H]
\vspace*{-2mm}
\centering
\includegraphics[scale=0.7]{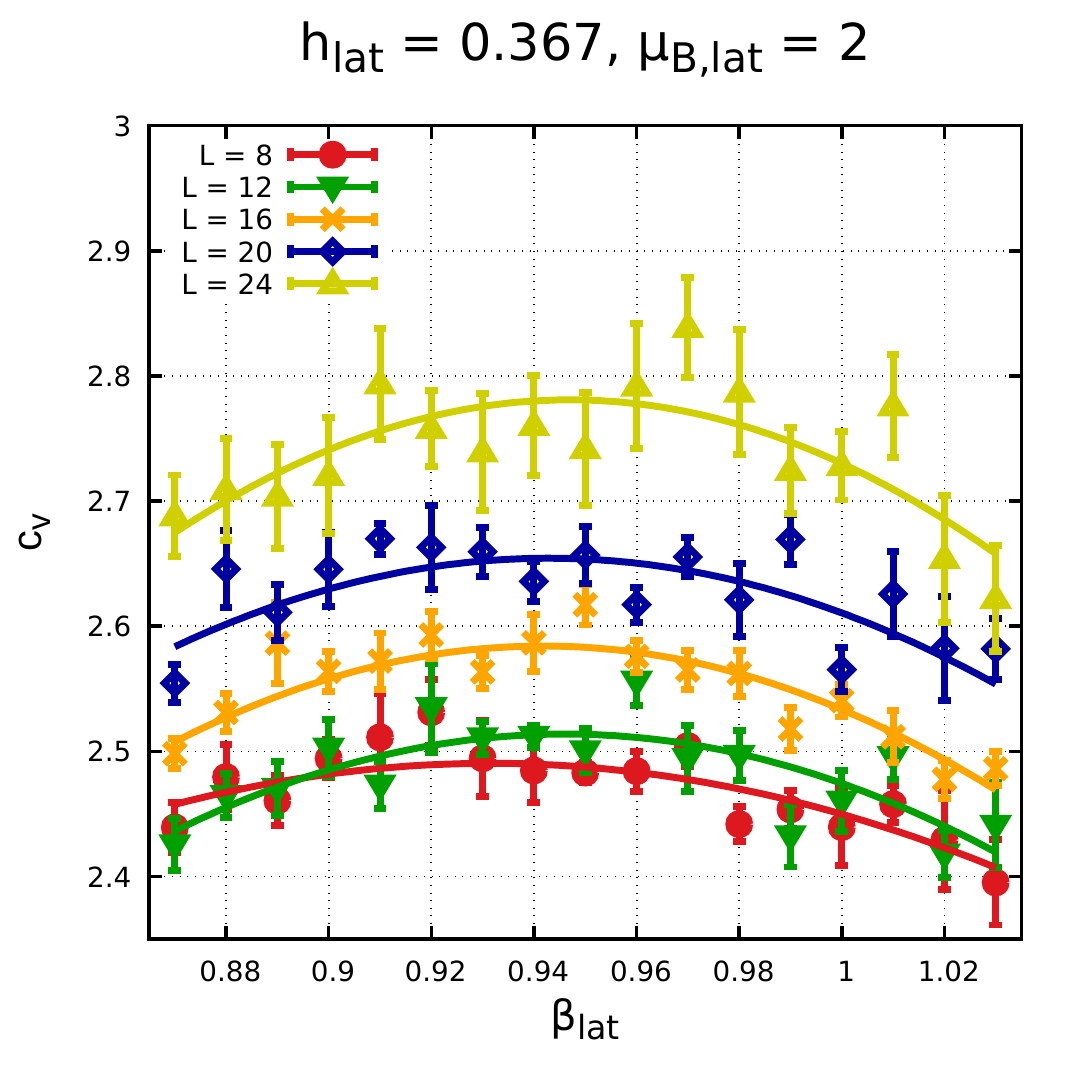}
\caption{The specific heat $c_{V}$ at $\mu_{B{\rm lat}} =2$ as a function
  of $\beta_{\rm lat}$, in various volumes. The peaks are smeared out due
  to $h_{\rm lat} > 0$.
  We localize the maxima $\beta_{\rm max,lat}$ by Gaussian fits.}
\vspace*{-0.2cm}
\label{cVmu2m}
\end{figure}

Regarding the phase diagram, we rely on the second derivatives of $F$
which we already considered in the chiral case: the specific heat $c_{V}$
and the magnetic susceptibility $\chi_{m}$, given in eqs.\
(\ref{cVdef}) and (\ref{chimdef}). Figure \ref{cVmu2m} shows
$c_{V}$ at $\mu_{B,{\rm lat}}=2$: there are no clear peaks, unlike
Fig.\ \ref{cVplot}, but we can identify the maxima by Gaussian
fits (their uncertainty is estimated by the jackknife method).

\begin{figure}[H]
\centering
\includegraphics[scale=0.7]{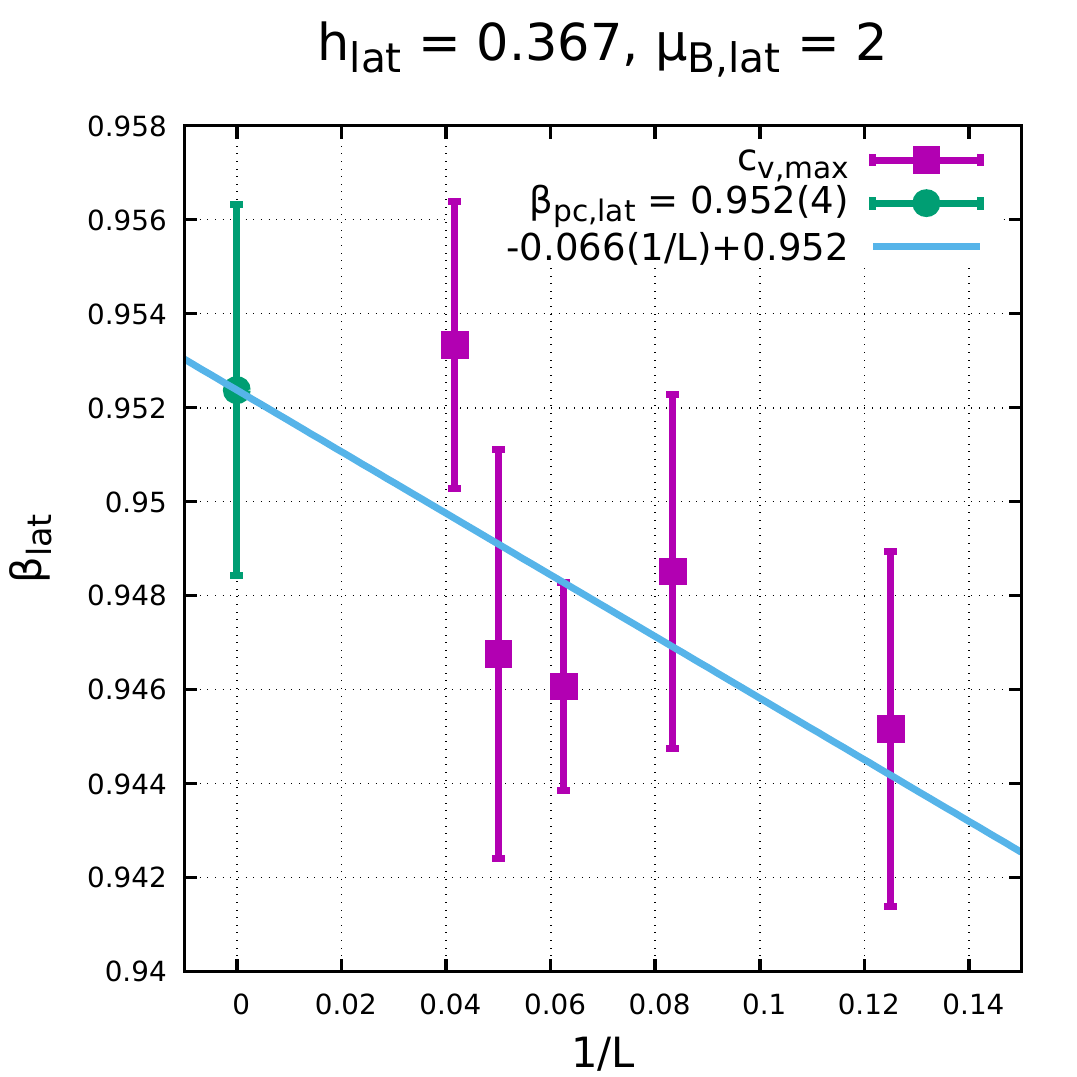}
\caption{Extrapolations of the maxima $\beta_{\rm max,lat}$ of $c_{V}$
  at $L= 8 \dots 24$ to the large-$L$ limit. This is carried out for
  each value of $\mu_{B,{\rm lat}}$; here we show $\mu_{B,{\rm lat}}=1$ and $2$
as examples.}
\vspace*{-0.2cm}
\label{cVextram}
\end{figure}

Figure \ref{cVextram} is devoted to the thermodynamic extrapolations
of these maxima at $\mu_{B,{\rm lat}} =1$ and $2$, which lead to
\bea
\beta_{\rm pc,lat}(\mu_{B,{\rm lat}} =1) &=& 0.956(6) \ , \nn \\
\beta_{\rm pc,lat}(\mu_{B,{\rm lat}} =2) &=& 0.952(4) \ .
\eea

\begin{figure}[H]
\centering
\includegraphics[scale=0.7]{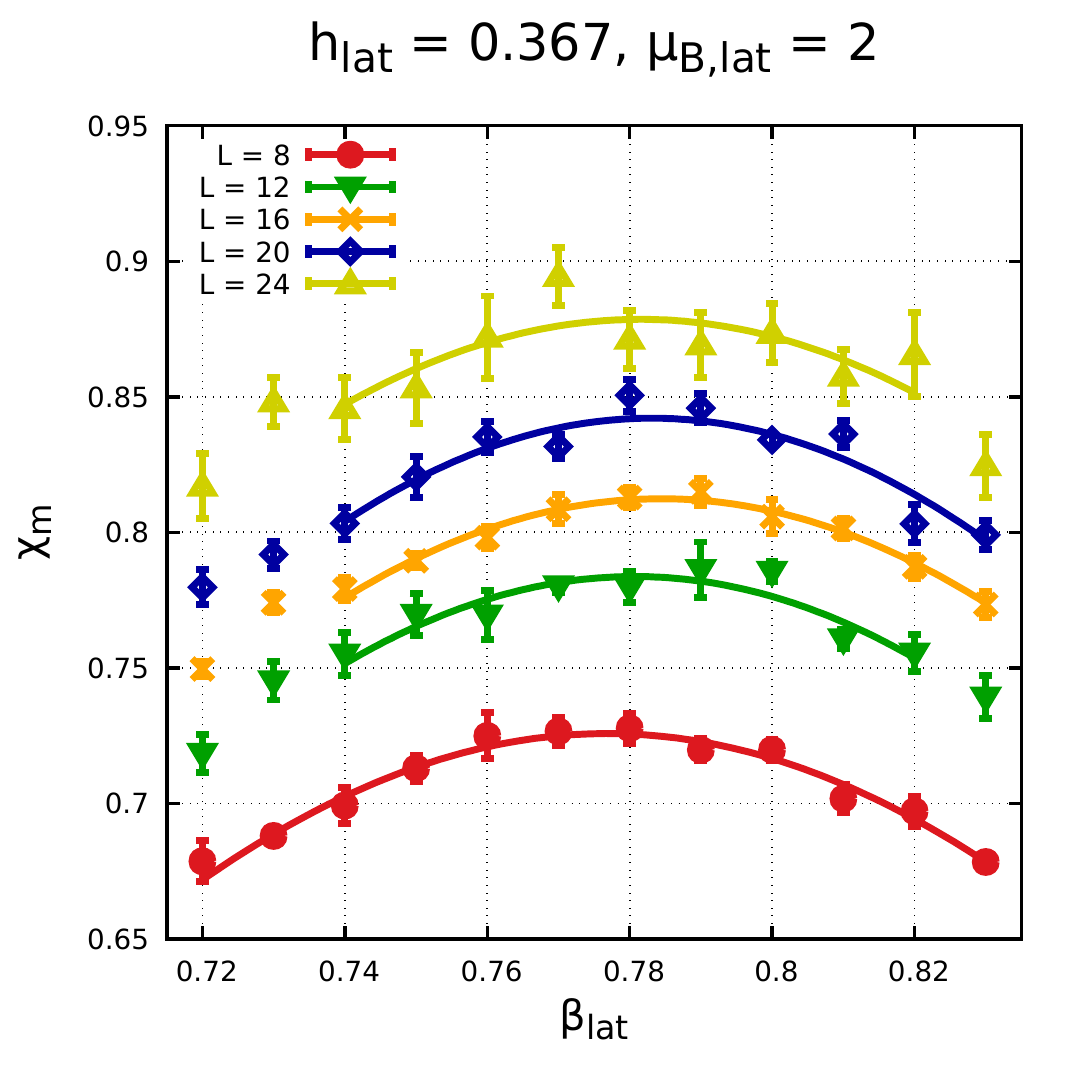}
\caption{The magnetic susceptibility $\chi_{m}$ at $\mu_{B,{\rm lat}} =2$
  as a function of $\beta_{\rm lat}$, in various volumes.
  Again we identify the maxima $\beta_{\rm max,lat}$ by Gaussian fits.}
\vspace*{-0.2cm}
\label{chimmu2m}
\end{figure}

Similarly, Fig.\ \ref{chimmu2m} shows $\chi_{m}$ at $\mu_{B,{\rm lat}} =2$.
Again there are no sharp peaks, but here the Gaussian fits work well
and provide another criterion for the pseudo-critical values
$\beta_{\rm pc,lat}$. We see that these values are well below the ones
obtained from $c_{V}$ in Fig.\ \ref{cVmu2m}.

\begin{figure}[H]
\centering
\includegraphics[scale=0.7]{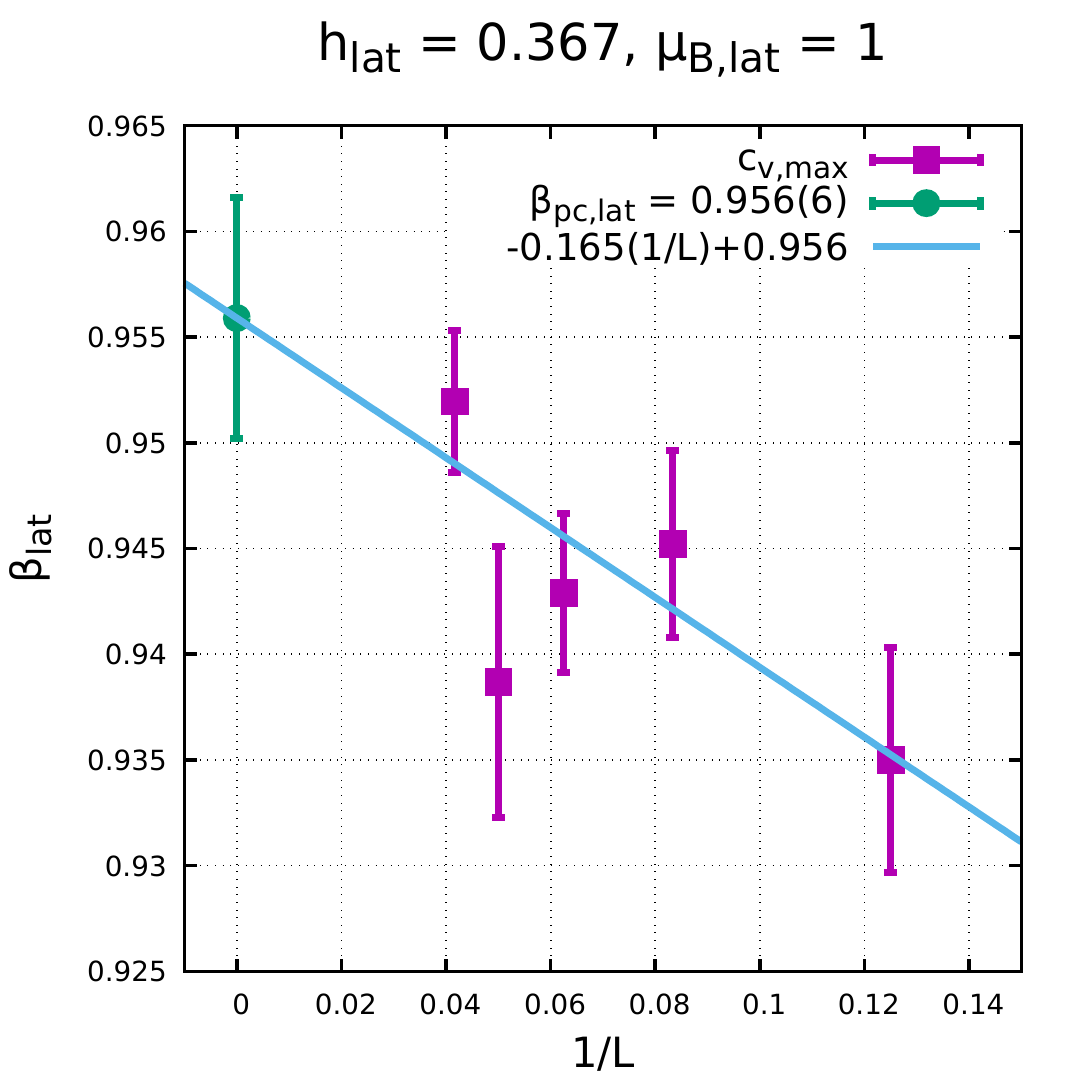}
\end{figure}

\begin{figure}[H]
\centering
\includegraphics[scale=0.7]{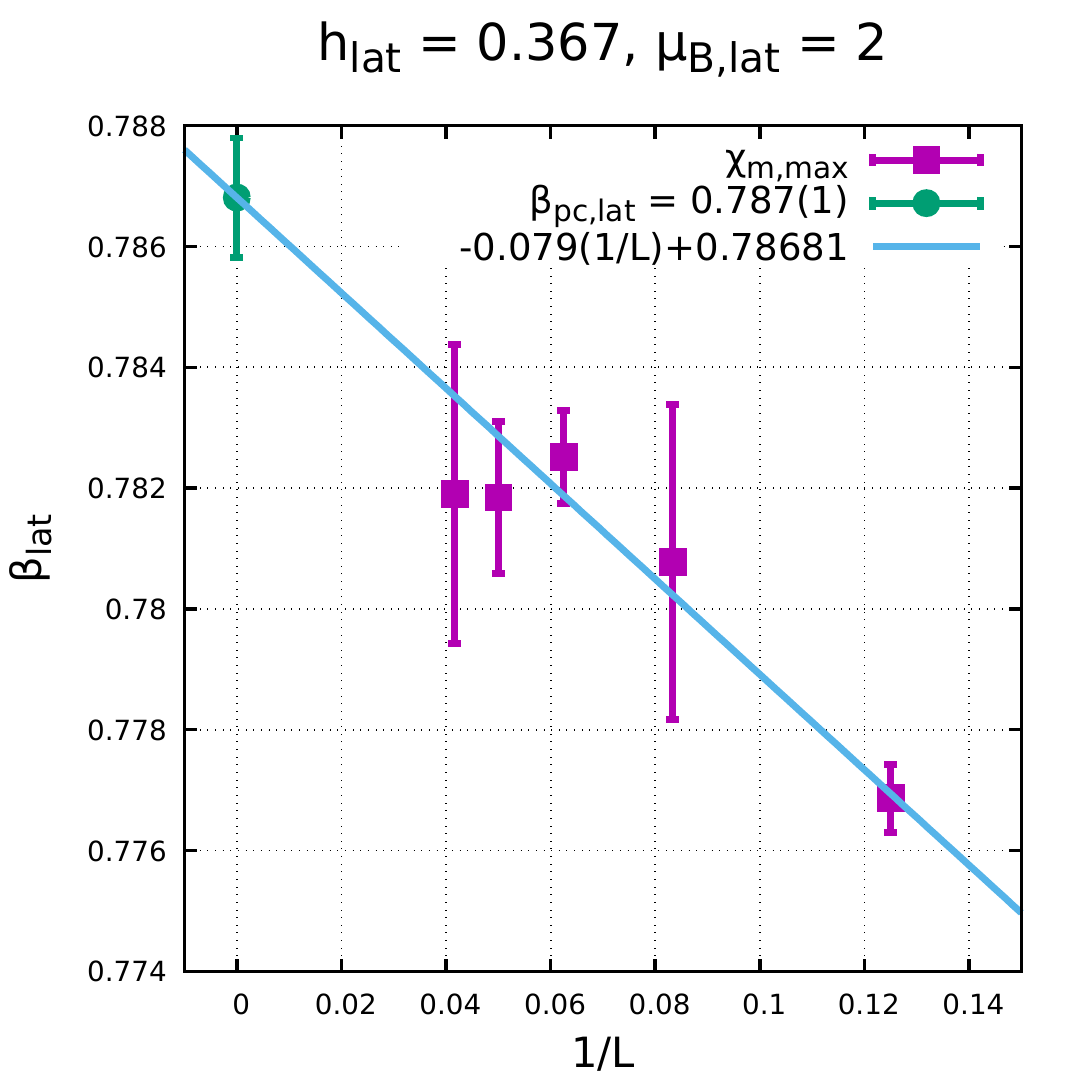}
\caption{Extrapolations of the maxima $\beta_{\rm max,lat}$ of $\chi_{m}$
  at $L= 8 \dots 24$ to the large-$L$ limit. We show the examples
  at $\mu_{B,{\rm lat}}=1$ and $2$.}
\vspace*{-0.2cm}
\label{chimextram}
\end{figure}

Figure \ref{chimextram} illustrates the large-$L$ extrapolations of these
maxima. In these examples, we obtain
\bea
\beta_{\rm pc,lat}(\mu_{B,{\rm lat}} =1) &=& 0.739(1) \ , \nn \\
\beta_{\rm pc,lat}(\mu_{B,{\rm lat}} =2) &=& 0.787(1) \ .
\eea

In Figs.\ \ref{cVextram} and \ref{chimextram} we see slopes which
are significantly driven by the result at $L=8$ --- an effect, which
we would like to overcome.
We should soon have results at $L>24$, which will enable sensible
fits excluding $L=8$. This will improve the validity of the
$\beta_{\rm pc,lat}$ values, although they will change most likely
just at percent-level.

\begin{figure}[H]
\centering
\includegraphics[scale=0.7]{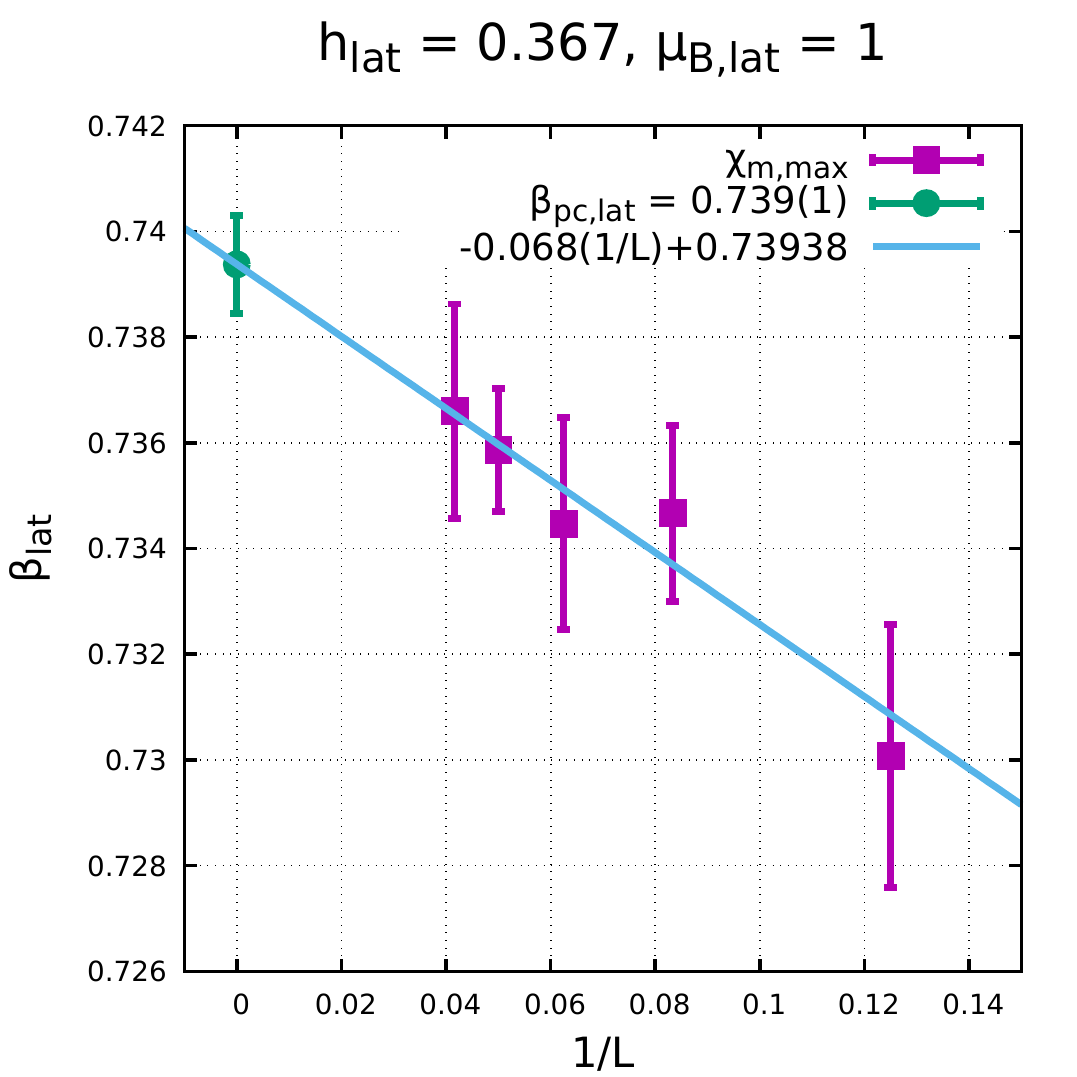}
\end{figure}
\vspace*{-10mm}

\begin{figure}[H]
\vspace*{-4.5cm}
\hspace*{-12mm}
\includegraphics[scale=0.545]{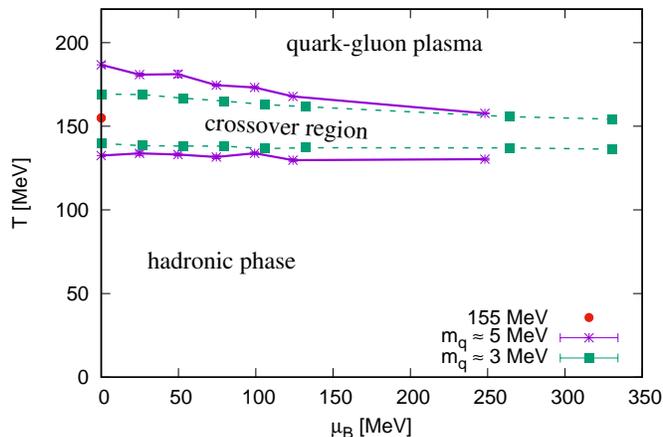}
\vspace*{-2cm}
\caption{Conjectured phase diagram of 2-flavor QCD with degenerate quark
  masses corresponding to $m_{u}$ or to $m_{d}$.}
\vspace*{-2mm}
\label{phasediagrammassive}
\end{figure}

As another working hypothesis, we interpret the large-$L$ extrapolated
results for $\beta_{\rm pc,lat}$ based on $c_{V}$ and on $\chi_{m}$ as
boundaries of the crossover interval. We convert $\mu_{B,{\rm lat}}$ and
$T_{\rm lat} = 1/\beta_{\rm lat}$ to physical units, as described in the
beginning of this section, and include the corresponding results that
we obtained at $h_{\rm lat}=0.14$, including $\mu_{B,{\rm lat}}=2.5$.
This leads to our conjectured phase diagram
in the massive case, which we display in Fig.\ \ref{phasediagrammassive}.
We recall that $h_{\rm lat} =0.14$ and $0.367$ (roughly) correspond to the
physical quark masses $m_{u}$ and $m_{d}$.

In contrast to the chiral phase diagram in Fig.\ \ref{phasediachiral},
we see only a weak trend of the crossover interval to bend down to lower
temperatures as $\mu_{B}$ increases up to $\approx 300~{\rm MeV}$.
On the other hand, just as in the chiral limit, we did not
encounter the notorious CEP in the range that we explored so far.

\section{Summary and conclusions}

We presented a study of the O(4) non-linear $\sigma$-model.
We have good reasons to assume this model to be in the same universality
class as 2-flavor QCD in the chiral limit, because the spontaneous
symmetry breaking patterns coincide.\footnote{Due to the dimensions
  of the Lie groups, O($N$) models cannot cope with $N_{\rm f} > 2$
  flavors \cite{WB}, cf.\ footnote 1. It is a drawback of this
effective theory that we cannot include the $s$-quark or even
heavier quark flavors.}

We are interested in high temperatures, which we assume to be high
enough to justify dimensional reduction as a reasonable approximation.
This leads to the 3d O(4) model, which has topological charges. They
correspond to the baryon number, as Skryme already knew even before
QCD was established \cite{Skyrme}.\footnote{Skyrme stayed in 4 space-time
  dimensions, and added a 4-derivative term in order to stabilize the
  structures in the configurations, which include a 3d spatial
  instanton.}

In this sense, the model can be simulated with a baryon chemical
potential $\mu_{B,{\rm lat}}$, which corresponds to an imaginary vacuum
angle, without any sign problem. As a further advantage, we can
apply a powerful cluster algorithm.\\

In the chiral limit, we followed the critical line up to
$\mu_{B} \simeq 309~{\rm MeV}$, $T_{\rm c} \simeq 106~{\rm MeV}$.
(We converted lattice units to physical units by referring to
the critical temperature at $\mu_{B}=0$.)

The result is shown in Fig.\ \ref{phasediachiral}.
The line for $T_{\rm c}(\mu_{B})$ decreases monotonically, in agreement
with other conjectures in the literature; this is the generally expected
behavior. As far as we could follow this line, we did not find a
Critical Endpoint (CEP), but there are hints for it to be near
the final point included in our study.\\

We also investigated the massive case, with degenerate
quark masses $m_{q}$, which approximately correspond either
to $m_{u}$ or to $m_{d}$ (this identification involves the chiral
condensate, in addition to the pseudo-critical temperature $T_{\rm pc}$).
Here, we identified an interval for the crossover temperature, based
on the maxima of two observables, which are given by second derivatives
of the free energy $F$ (in the chiral limit, they
detect the critical temperature).

We monitored this crossover interval up to $\mu_{B} \approx 300~{\rm MeV}$.
It is rather broad, see Fig.\ \ref{phasediagrammassive}, with only a
minor trend towards lower $T_{\rm pc}$ as $\mu_{B}$ increases.
Again we could not find a CEP.

An overview over predictions for the CEP temperature and baryonic chemical
potential is given in Ref.\ \cite{Lat21}, and compared to the bounds based
on our conjecture.

If we manage to extend the numerical study of this effective theory
to larger values of $\mu_{B}$ and eventually find a CEP, we could also
explore phases at even higher $\mu_{B}$ that the literature speculates
about.\\

\noindent
{\bf Acknowledgments:} We are indebted to Arturo Fern\'{a}ndez T\'{e}llez
and Miguel \'{A}ngel Nava Blanco for their contributions to this project
at an early stage. We thank Uwe-Jens Wiese for instructive discussions,
and Philippe de Forcrand and Christian Hoelbling for interesting comments.
This work was presented by ELC and JAGH at the
{\it XXXV Reuni\'{o}n Anual de la Divisi\'{o}n de Part\'{\i}culas y
Campos} of the {\it Sociedad Mexicana de F\'{\i}sica}.
The simulations were performed on the cluster of the Instituto
de Ciencias Nucleares; we thank Luciano D\'{\i}az Gonz\'{a}lez
for technical assistance.
We acknowledge support by UNAM-DGAPA through PAPIIT project IG100219,
``Exploraci\'{o}n te\'{o}rica y experimental del diagrama de fase de
la cromodin\'{a}mica cu\'{a}ntica'', and by the Consejo Nacional de
Ciencia y Tecnolog\'{\i}a (CONACYT).

\end{multicols}
\medline
\begin{multicols}{2}

\end{multicols}
\end{document}